\documentstyle[12pt,epsf]{article}

\font\sf=cmss10                    
\def\half{\frac{1}{2}}
\def\ra{\rightarrow}

\def\Tr{\rm Tr\ }
\def\ch#1#2{({#1 \atop #2 })}
\newcommand{\mathbold}[1]{\mbox{\boldmath $\bf#1$}}
\def\A{{\cal A}}  
\def\mJ{\mathbold{J}}
\def\mA{\mathbold{A}}
\def\mB{\mathbold{B}}
\def\mC{\mathbold{C}}
\def\mX{\mathbold{X}}
\def\mG{\mathbold{G}}
\def\mx{\mathbold{x}}
\def\mZ{\mathbold{Z}}
\def\mz{\mathbold{z}}
\def\malpha{\mathbold{\alpha}}
\def\momega{\mathbold{\omega}}

\def\bbbz{{\sf Z\!\!\!Z}}
\def\sl2z{SL(2,\bbbz)}

\def\ll{\lambda \cdot \lambda}

\def\LL{\Lambda \cdot \Lambda}
\def\zz{{\bf z}}
\def\z0{{\bf z_0}}
\def\gg{{\cal G}}

\def\bbbp{\,{\hbox{P}\!\!\!\!\!\hbox{I}\,\,\,}}

\newcommand{\be}{\begin{equation}}
\newcommand{\ee}{\end{equation}}
\newcommand{\bea}{\begin{eqnarray}}
\newcommand{\eea}{\end{eqnarray}}
\newcommand{\nn}{\nonumber}
\newcommand{\bean}{\begin{eqnarray*}}
\newcommand{\eean}{\end{eqnarray*}}
\newcommand{\myref}[1]{(\ref{#1})}
\newcommand{\secref}[1]{sec.~\protect\ref{#1}}
\newcommand{\figref}[1]{Fig.~\protect\ref{#1}}
\newcommand{\mat}[4]{\left(\begin{array}{cc} #1 & #2 \\ #3 & #4
\end{array}\right)}
\newcommand{\vect}[2]{({#1\atop #2})}
\newcommand{\unit}{1\!\!1}
\newcommand{\mtimes}{\!\times\!}
\newcommand{\mQ}{\ell}
\newcommand{\genh}{\gg^{enh}}
\newcommand{\Genh}{{\bf G}^{enh}}
\newcommand{\comment}[1]{}

\newcommand{\onefigure}[2]{\begin{figure}[htbp]
         \caption{#2\label{#1}(#1)}
         \end{figure}}
\renewcommand{\onefigure}[2]{\begin{figure}[htbp]
         \begin{center}\leavevmode\epsfbox{#1.eps}\end{center}
         \caption{#2\label{#1}}
         \end{figure}}

\begin{document}

\setlength{\baselineskip}{14pt}
\setlength{\parskip}{1.35ex}
\setlength{\parindent}{0em}
\renewcommand{\theequation}{\thesection.\arabic{equation}}

\noindent

\thispagestyle{empty}
{\flushright{\small MIT-CTP-2804\\hep-th/9812028\\}}

\vspace{.3in}
\begin{center}\Large {\bf Uncovering the Symmetries on [p,q]
7-branes: \\
Beyond the Kodaira Classification}
\end{center}

\vspace{.1in}
\begin{center}
{\large Oliver DeWolfe, Tam\'as Hauer, Amer Iqbal and Barton Zwiebach}

\vspace{.1in}
{ {\it Center for Theoretical Physics,\\
Laboratory for Nuclear Science,\\
Department of Physics\\
Massachusetts Institute of Technology\\
Cambridge, Massachusetts 02139, U.S.A.}}
\vspace{.2in}

E-mail: {\tt odewolfe,hauer,iqbal@ctp.mit.edu, zwiebach@irene.mit.edu}
\end{center}
\begin{center}December 1998\end{center}

\vspace{0.1in}

\begin{abstract}
We begin a classification of the symmetry algebras arising on
configurations of type IIB $[p,q]$ 7-branes.  These include not just
the Kodaira symmetries that occur when branes coalesce into a
singularity, but also algebras associated to other physically
interesting brane configurations that cannot be collapsed.  We
demonstrate how the monodromy around the 7-branes essentially
determines the algebra, and thus 7-brane gauge symmetries are
classified by conjugacy classes of the modular group $\sl2z$.  Through
a classic map between the modular group and binary quadratic forms,
the monodromy fixes the asymptotic charge form which determines the
representations of the various $(p,q)$ dyons in probe D3-brane
theories.  This quadratic form also controls the change in the algebra
during transitions between different brane configurations.  We give a
unified description of the brane configurations extending the $D_N$,
$E_N$ and Argyres-Douglas $H_N$ series beyond the Kodaira cases.  We
anticipate the appearance of affine and indefinite
infinite-dimensional algebras, which we explore in a sequel paper.
\end{abstract}

\newpage
\section{Introduction}

Over the last couple of years F-theory \cite{vafa} compactifications
on an elliptic K3 have been studied explicitly as type IIB
compactifications involving of $[p,q]$ 7-branes. In particular, an
elliptic K3 with a Kodaira singularity maps to a 7-brane configuration
where the branes have coalesced to form the singularity. It has become
clear, however, that the Kodaira singularities do not exhaust the set
of 7-branes configurations relevant to field theory applications.
There exist 7-brane configurations that cannot be collapsed, but
nevertheless provide backgrounds for interesting D3-brane probe
theories.  The most familiar examples of such theories are the ${\cal
N} = 2$, $D=4$ Seiberg-Witten theories \cite{seibergwitten} with $N_f
\leq 3$ flavors.  These can be realized in the presence of background
7-brane configurations of $D_{ N_f}$ global symmetry type, but the
branes cannot be collapsed since Kodaira singularities of $D_N$ type
only appear for $N \geq 4$.  Other examples include the
four-dimensional theories with $E_N$ global symmetry, where any value
$N \geq 0$ arises from a possible 7-brane background configuration,
but only for $N = 5,6,7,8$ the 7-branes can be collapsed to a
singularity.

These applications, as well as others, have led us to consider
enumerating all the possible 7-brane configurations and their
corresponding algebras, whether or not there is an associated
singularity.  Finite Lie algebras arising on 7-brane configurations
were explored in
\mbox{\cite{johansen,GZ,mgthbz,imamura,dewolfezwiebach}}, and it was
shown in detail how $(p,q)$ strings and string junctions stretched
between the 7-branes correspond to vector bosons of the
eight-dimensional gauge theory.  In M-theory these junctions lift to
M2-branes embedded in K3, and the requirement that the states be BPS
specifies that these branes must be holomorphically embedded, {\it
i.e.}  they are wrapped on holomorphic 2-cycles of the K3
\cite{nekrasov}.  An inner product on junction space is induced from
the intersection of the associated 2-cycles, and the Cartan matrix of
the Lie algebra appears as the intersection matrix of a basis of
string junctions ending on the 7-brane configuration.  The algebra is
then realized on the branes through the composition of junctions. A
key result of \cite{dewolfezwiebach} was the expression of the
self-intersection of a junction emerging from a 7-brane configuration
with ``asymptotic'' charges $(p,q)$,
\begin{equation}
\label{thefirst}
\mJ^2 = - \ll + f\,(p,q) \,,
\end{equation}
where $\lambda$ is the Lie algebra weight vector associated to the
junction and $f(p,q)$ is a quadratic form determined by the 7-brane
configuration.

The requirement of holomorphy implies a constraint on the
self-intersection of a junction, and this was exploited in \cite{DHIZ}
to determine the possible BPS junctions on theories with a single
probe D3-brane in a 7-brane background.  The BPS spectra of $SU(2)$
Seiberg-Witten theory for $N_f \leq 4$ was reproduced exactly using
only this constraint.  (A related constraint was used to obtain the
$N_f=0$ case in \cite{nekrasov}.)
Results were derived for the largely unknown
spectra of the four-dimensional theories with $E_N$ global
symmetry. Much about their spectra is controlled by the quadratic form
$f(p,q)$ appearing in (\ref{thefirst}).  The computation of
self-intersection for junctions for the case of 7-brane backgrounds
and multiple D3-branes was given in \cite{iqbal}.

In this paper we expand on and systematize this previous work.  We
begin a classification of all the possible configurations of 7-branes,
and their associated algebras.  We find that the monodromy matrix
associated to a set of 7-branes essentially specifies which Lie
algebra is realized. More precisely, in all cases we have considered,
the $\sl2z$ conjugacy class of the monodromy, together with the number
of branes and an integer $\mQ$ characterizing the possible asymptotic
charges of junctions, determine the algebra uniquely.

Thus we find an elegant organization of all possible Lie algebras on
7-branes based on conjugacy classes of $\sl2z$ matrices. The
classification of such conjugacy classes is a well-known mathematical
problem closely related to the classification of equivalence classes
of binary quadratic forms. This relation is implemented by a map
associating binary quadratic forms to elements of $\sl2z$.  According
to the value of its trace, an $\sl2z$ matrix is either elliptic,
parabolic or hyperbolic. We find that the finite algebras
corresponding to Kodaira singularities exhaust the elliptic classes
and fill some of the parabolic classes, while the other
non-collapsible finite algebra configurations are in parabolic classes
and hyperbolic classes of negative trace.

Although the conjugacy classes of $\sl2z$ matrices are the primary
organizational tool, the classification also requires the number of
branes and the integer $\mQ$, and perhaps other data we have not yet
encountered. Specifying the number of branes is necessary because
there exist configurations of twelve branes with unit monodromy, which
nonetheless change the algebra of a given configuration
considerably. When different from one, $\mQ$ indicates that not all
possible asymptotic charges can appear on junctions.  We have found
configurations that have the same monodromy and the same number of
branes but give different algebras; in such cases $\mQ$ differs.
Similarly, configurations with the same algebras and number of branes
will have inequivalent monodromies if $\mQ$ is distinct.

We also show that the map from $\sl2z$ to quadratic forms, applied to
the monodromy matrix of the brane configuration, gives us the
corresponding asymptotic charge quadratic form $f(p,q)$ appearing in
(\ref{thefirst}).  We explore systematically how this quadratic form
controls the change in the Lie algebra when a new brane is introduced.
We find that $f(p,q)$, where $f$ is the asymptotic charge form for the
original configuration and $[p,q]$ are the charges of the new brane,
precisely encodes the type of enhancement that takes place.  When
$f(p,q) < -1$ the original algebra is only supplemented by a new
$u(1)$, while for precisely $f(p,q)=-1$ an additional $A_1$ appears.
For values $-1 < f(p,q) < 1$, the original algebra enhances to some
other finite algebra with rank greater by one.

We also present a simple and unified description of the ${\bf D_N}$,
${\bf E_N}$ and ${\bf H_N}$ series of configurations, which realize
$D_N$, $E_N$ and $A_N$ algebras respectively.  These configurations
can collapse into Kodaira singularities for ${\bf D_N}$, $N \geq 4$,
for ${\bf E_5}$, ${\bf E_6}$, ${\bf E_7}$ and ${\bf E_8}$, and finally
for ${\bf H_0}$, ${\bf H_1}$ and ${\bf H_2}$, the configurations
associated to Argyres-Douglas points.  The configurations ${\bf D_5}$
and ${\bf E_5}$ both realize a $D_5$ algebra, and as an example we
show explicitly how one configuration can be transformed into the
other by a global $\sl2z$ transformation and by relocation of branch
cuts.

Once one considers brane configurations that are not collapsible,
infinite-dimensional algebras appear naturally.  This was seen in
\cite{dewolfe}, where it was shown that affine Lie algebras arise
whenever the junction intersection form produces an affine Cartan
matrix.  We discuss how these and other infinite-dimensional Lie
algebras appear in our classification, filling many of the remaining
$\sl2z$ conjugacy classes and corresponding to brane transitions with
$f(p,q) \geq 1$, in the sequel paper \cite{infinite}.

In the present paper we also take the opportunity to discuss
explicitly a basic issue that is seldom brought into the open. We
explain in general terms how a Lie algebra arises from BPS junctions
(or holomorphic 2-cycles). The conventional wisdom is that once the
intersection matrix of a set of basis junctions gives the Cartan
matrix of a Lie algebra, that algebra is realized. We give evidence
for this idea by showing that when a set of junctions represent {\it
simple roots}, namely, their intersections define a Cartan matrix, the
constraints of holomorphicity applied to junctions built from these
basis junctions naturally take the form of Serre relations.  These
relations indeed constrain the combinations of simple roots that
define allowed roots in the Chevalley-Serre construction of a Lie
algebra starting from a Cartan matrix. We also indicate what aspects
of the construction survive when the algebra to be identified has no
Cartan matrix, or when we deal with nontrivial infinite dimensional
algebras.

This paper is organized as follows. In section 2 we show how to
compute efficiently monodromies and their traces. We define carefully
the notion of equivalent brane configurations, illustrating it with
examples.  We also consider explicitly the classification of
configurations with two 7-branes.  In section 3 we begin by
explaining how a Lie algebra arises from junctions. We then show how
the monodromy of a configuration determines the associated charge
quadratic form, and prove a relation between the trace of the
monodromy and the determinant of the Cartan matrix associated to the
algebra arising from the brane configuration. In section 4 we consider
the unified presentation and extension of the main Kodaira series, as
well as generalizations thereof.  In sections 5 and 6 we discuss
transitions between brane configurations, limiting ourselves to the
case when the resulting algebra is finite. In section 7 we collect the
results relevant to the general classification of brane
configurations, anticipating some of the results to be explained in
\cite{infinite}.

\section{7-Brane Configurations and Monodromies}
\label{monodromy}

In this section we first discuss a few simple computations involving
branes and their monodromies. These include $\sl2z$ conjugation,
computations of traces, and transformations induced by moving branch
cuts. We then explain in detail our notion of equivalent brane
configurations. This notion is illustrated with some examples. One
noteworthy example also illustrates an exception, namely that in the
exceptional series $E_N$ there are two inequivalent brane
configurations associated to $E_1$. Finally, we discuss a
classification of brane configurations involving two 7-branes.

\subsection{Monodromy matrices and moving branch cuts}

Throughout the paper we shall be considering configurations of IIB
7-branes, filling (7+1) dimensions and pointlike in the remaining two,
which we take to be the complex plane.  These 7-branes are magnetic
sources for the complex dilaton-axion scalar $\tau = \chi + i e^{-
\phi}$.  This scalar experiences a monodromy around each 7-brane,
which we account for by introducing a branch cut associated to each
7-brane.  Following the conventions of \cite{mgthbz,dewolfezwiebach},
a generic background is specified by listing the 7-branes in the order
in which their branch cuts are crossed when encircling them in a
counterclockwise direction.

A 7-brane is labeled by two relatively prime integers $(p,q)$ up to a
sign; a $[p,q]$ 7-brane is the same object as a $[-p,-q]$ 7-brane. It
is convenient to place the branes in a {\em canonical presentation},
locating them along the real axis, ordered from left to right and with
the cuts going downwards. When crossing the cut of an individual
$[p,q]$-brane, $\tau$ transforms with the $\sl2z$ monodromy matrix
$K_{[p,q]}$ \cite{vgs}. It is convenient to unify the two charges in a
vector ${\bf z}=\ch{p}{q}$, in terms of which $K_{\bf z}$ is given as
\bea K_{\bf z} = \unit + {\bf z}{\bf
z}^T S &=& \mat{1+pq}{-p^2}{q^2}{1-pq} \,,
\label{Kdef}
\end{eqnarray}
where $S  \equiv \pmatrix{0 &-1 \cr 1 & 0}$. Under the operation of
global $\sl2z$ conjugation with an element $g$ one finds
\begin{equation}
\label{congg}
g\, K_\zz\, g^{-1} = K_{g\,\zz}\,.
\end{equation}

Following the conventions stated above, the total monodromy around the
brane configuration
$\mX_1 \ldots \mX_{n-1} \mX_n \,,$
is given by
\begin{equation}
K=K_{\zz_n}K_{\zz_{n-1}}\ldots K_{\zz_1} \,,
\end{equation}
and will be of vital importance to us.  The labeling of branes
actually depends on the placement of branch cuts.  We can move the
branch cut of one 7-brane $\mX_{\zz_1}$ across another 7-brane
$\mX_{\zz_2}$, thus changing the latter to $\mX_{\zz'_2}$ and
exchanging their order in the canonical presentation, or we can move
the cut of $\mX_{{\zz_2}}$ across $\mX_{\zz_1}$. This was explained in
\cite{mgthbz},
and the result can be written as follows:
\begin{eqnarray}
\label{crosstrans}
{\bf X}_{\displaystyle\zz_1}  {\bf X}_{ \displaystyle\zz_2}  &=&
{\bf X}_{\displaystyle\zz_2 }\,
{\bf X}_{\,\displaystyle (\zz_1 + (\zz_1 \times \zz_2)\,
\zz_2\, )} \nonumber \\
&=& {\bf X}_{ \,\displaystyle
(\zz_2 + (\zz_1 \times \zz_2)\, \zz_1\, )}\,
{\bf X}_{\displaystyle\zz_1 }\,,
\end{eqnarray}
where we have defined
\begin{equation}
\label{introdet}
{\bf z}_1 \mtimes {\bf z}_2 \equiv -{\bf z}_1^TS\, {\bf z}_2 =
{\bf z}_2^TS\, {\bf z}_1 =
\hbox{det}\, \pmatrix{ p_1 & p_2 \cr q_1 & q_2}\,.
\ee
Equation (\ref{crosstrans}) indicates the fixed brane acquires an
extra charge equal to the charge of the moving brane times the
determinant of the relative charges.

For a given brane configuration the trace of the associated monodromy
is an $\sl2z$ invariant.  Given the configuration ${\mX_1 \mX_2 \cdots
\mX_n}$ with monodromy $K = K_n \cdots K_2 K_1$ the trace of $K$ is
calculated using \myref{Kdef}. One finds:
\begin{equation}
\label{threemon}
\hbox{Tr}\, K = 2 + \sum_{k=2}^n \,\,\,\,\,\,\,\,\!\!\!\!\!\!
\sum_{i_1<i_2 \ldots < i_k}
({\bf z}_{i_1}\mtimes{\bf z}_{i_2} )
({\bf z}_{i_2}\mtimes{\bf z}_{i_3} )
\ldots ({\bf z}_{i_k}\mtimes{\bf z}_{i_1} ).
\end{equation}

Another useful $\sl2z$ invariant of a brane configuration $\mX_1 \mX_2
\ldots \mX_n$ is the positive integer $\ell$, the greatest common
divisor of all non-vanishing pairwise determinants:
\begin{equation}
\label{l}
\mQ \equiv \left\{\begin{array}{ll}
\gcd\{{\bf z}_i\times{\bf z}_j, \mbox{for all $i,j$}\;
\}\,  &
\mbox{for mutually nonlocal branes}
\\
0 &
\mbox{for mutually local branes}.
\end{array}\right.
\end{equation}
For reasons that will be explained at the beginning of sect.~3, we
sometimes
call $\ell$ the asymptotic charge invariant.  
It is manifest that
$\mQ$ is invariant under global $\sl2z$ transformations, and it can be
shown it also does not change when branch cuts are relocated.

In this paper we continue to label some useful branes in the way we
did in previous works. We take $\mA = {\bf [1,0]}$ ($K_{\bf A} = T^{-1}$), $\mB =
{\bf [1,-1]}$ ($ K_{\bf B} = ST^2$), and $\mC = {\bf [1,1]}$ ($K_{\bf C} = T^2S$).  
Branes of
other charges $\zz$ will be denoted by $\mX_{\zz}$.

\subsection{Equivalence classes of brane configurations}

The classification of the algebras that can arise on configurations of
7-branes has to take into account equivalence transformations between
different configurations. We define:

{\it Two 7-brane configurations will be said to be equivalent
(indicated as $\cong$) if they have the same number of branes and
their canonical presentations can be matched brane by brane using the
operations of overall $\sl2z$ conjugation and relocation of branch
cuts.}

Global $\sl2z$ conjugation changes some brane labels and may or may
not change the overall monodromy.  Due to the $\sl2z$ symmetry of type
IIB string theory, it does not change the physics and is therefore an
equivalence transformation when applied to a complete configuration.
Moving branch cuts, as reviewed in the previous subsection, does not
change the overall monodromy, only the labels of the individual
7-branes. Two brane configurations related only by relocation of
branch cuts will be said to be {\it equal}.

Brane configurations can only be equivalent if their monodromies are
conjugate in $\sl2z$.  In addition, they must have the same value of
the invariant $\ell$. In all the examples we have studied, we have
found that two configurations with conjugate monodromies, equal values
for $\ell$ and the same number of branes are equivalent. Although we
have no general proof, we conjecture that

{\it Conjecture: Inequivalent 7-brane configurations are classified by
monodromy,
number of branes and the asymptotic charge invariant $\ell$.}

Classifying equivalence classes of monodromies in $\sl2z$ is a
well-studied but complicated problem. It is useful to consider the
trace of the monodromy, which is an $\sl2z$ invariant.  Two
monodromies with different trace are necessarily inequivalent, but if
their traces agree they still need not be equivalent.  Indeed, there
are inequivalent conjugacy classes in $\sl2z$ with the same trace.

In the remainder of this section, we will work through an example of
brane configurations that are equivalent, and we prove this completely
by mapping the brane configurations into each other explicitly.
Another similar example is given in the sequel paper \cite{infinite}.
Following this we present two different series realizing the $E_N$
algebras, and prove the equivalence of all pairs in the two series
except for one pair, having to do with $E_1$.  This exception
illustrates how differing $\ell$ can render configurations
inequivalent.

\noindent
{\bf Equivalence of two realizations of so(10)}.
Consider the conventional ${\bf D_5}= \mA^5\mB\mC$ configuration of
branes
known to give the $D_5 = so(10)$ algebra. On the other
hand the $E_5$ algebra of the exceptional series is isomorphic to
$D_5$. Since $E_6$, $E_7$ and $E_8$ are realized on ${\bf E_6} =
\mA^5\mB\mC\mC$, ${\bf E_7} = \mA^6\mB\mC\mC$, and ${\bf E_8} =
\mA^7\mB\mC\mC$, it is natural to ask whether the brane configuration
${\bf E_5} = \mA^4\mB\mC\mC$ gives an equivalent construction of
$so(10)$.  The answer is yes. To show this we first confirm that the
two monodromies are $\sl2z$ conjugate
\begin{equation}
K ({\bf E_5}) = - K_{\bf [1,1]} =
\mat{-2}{1}{-1}{0} \sim \mat{-1}{1}{0}{-1} = - K_{\bf [1,0]} =
K ({\bf D_5}) \, .
\end{equation}
The $\sl2z$ transformation can be taken to be the transformation that
acting on the $[1,0]$ brane gives the $[1,1]$ brane
\begin{equation}
K_{\bf [1,1]} = g \,K_{\bf [1,0]}\,\,g^{-1} \,, \qquad
g = \pmatrix{1&0\cr 1&1},
\end{equation}
for example. This transformation turns $\mA^5\mB\mC$ into $\mC^5 \mA
\mX_{\bf [1,2]}$. This configuration must now be shown to be identical
to $\mA^4\mB\mC\mC$ by moving cuts.  By repeated use of
eqn.~(\ref{crosstrans}) we find the claimed equivalence:
\begin{eqnarray}
\mC^5 \mA \, \mX_{\bf [1,2]} &=& \mA \, (\mX_{\bf [0,1]})^5 \, \mX_{
\bf[1,2]} = \mA\, (\mX_{\bf [0,1]})^4 \, (\mX_{\bf [0,1]} \mX_{\bf
[1,2]})  = \mA\, (\mX_{\bf [0,1]})^4 \,\mC\, \mX_{\bf [0,1]}
\nonumber\\
&=& \mA\, (\mX_{\bf [0,1]})^2 \,\mC\,\mA^2\, \mX_{\bf [0,1]}  = \mA\,
(\mX_{\bf [0,1]})^2 \, \mA^2 \, \mB \, \mX_{\bf [0,1]} \nonumber \\
&=& \mA\, ((\mX_{\bf [0,1]})^2 \, \mA ) \, \mA \, (\mB \, \mX_{\bf
[0,1]}) = \mA^2 \mB^2 \mA^2 \mB \\
&=&  \mA^4 \, (\mX_{\bf [3,-1]})^2 \, \mB = \mA^4 \mB\mC\mC
\,.\nonumber
\end{eqnarray}
\noindent
{\bf Equivalence and inequivalence in the $E_N$-series}.  We now
examine two series of algebras which realize the $E_N$ algebras, ${\bf
E_N} ={\bf A}^{N-1}{\bf BCC}$ and ${\bf \tilde{E}_N} ={\bf A}^{N}{\bf
X}_{\bf [2,-1]}{\bf C}$.  The first is a familiar series \cite{johansen,
GZ}, while the second will be introduced and explained in section 4.

Note that the second series gives a definition for
${\bf E_0}$ as $\mX_{\bf [2,-1]} \mC$ while the first series does
not.
The two series are equivalent for \ $N \geq 2$. To prove this it is
enough to concentrate on the four rightmost branes:
\begin{eqnarray}
{\bf ABCC} = {\bf BX}_{\bf [0,1]}{\bf CC} = {\bf BAAX}_{\bf [0,1]} =
{\bf AAX}_{\bf [3,-1]}{\bf X}_{\bf [0,1]} \stackrel{K_A^{-1}}{\cong}
{\bf AAX}_{\bf [2,-1]}{\bf C},
\label{enequiv}
\end{eqnarray}
where the last step involves conjugation with $K_A$, which does not
affect the rest of the spectator {\bf A}-branes. The steps in
\myref{enequiv}, however, cannot be applied to the case of fewer then
four branes.  This means that ${\bf E_1}={\bf BCC}$ and ${\bf
\tilde{E}_1}={\bf AX_{\bf [2,-1]}}{\bf C}$ need not be equivalent.  In
fact, they are not.  We readily find from \myref{l} that $\ell ({\bf
E_1}) = 2$, while $\ell ({\bf \tilde{E}_1}) = 1$, thus guaranteeing
inequivalence.  Note that $\ell({\bf E_N}) = 1$ for $N \geq 2$.

\section{String Junctions, Lie Algebras and Quadratic Forms}

Thus far we have discussed only the $[p,q]$ 7-branes themselves.
Configurations of 7-branes support strings and string junctions
stretched in between them.  There exist BPS junctions realizing
the adjoint of the 7-brane algebra $\gg$, and in the case of $\gg$
finite they are gauge bosons living on the 7-brane worldvolume theory,
as we now review.

The labels $(p,q)$ are also associated to segments of strings
corresponding to bound states of fundamental and D-strings
\cite{witten/aharony/schwarz}.  In crossing the branch cut of an
$[r,s]$ 7-brane in the counterclockwise direction, the $(p,q)$ charges
of a string segment change as $\ch{p}{q} \ra K_{[r,s]} \,
\ch{p}{q}$. Under global $g \in \sl2z$ transformations all charges
carried by strings change as \mbox{$\ch{p}{q}\ra g\,\ch{p}{q}$}.  A
$(p,q)$ string can end only on a $[p,q]$ 7-brane, a statement that is
$\sl2z$-invariant as 7-brane labels $[r,s]$ change as $\ch{r}{s} \ra g
\, \ch{r}{s}$, as well as invariant under moving branch cuts.

A web of $(p,q)$-strings, here called a junction, is charged under the
$u(1)$ gauge field of a given 7-brane $\mX$ if the associated
invariant charge $Q_{\bf X}$ defined in \cite{dewolfezwiebach} is
non-vanishing.  This charge combines the contribution from crossing
the cut of $\mX$ and the contribution of string prongs emanating from
$\mX$ in a way that is invariant under Hanany-Witten transformations
\cite{hanany}.  We call two junctions equivalent if they are related
by junction transformations as defined in
\cite{dewolfezwiebach,hauer}; it was proven in \cite{hauer} that the
BPS representative of an equivalence class of junctions containing an
open string is unique. The algebraic properties of a given string
junction are entirely specified by the set of invariant charges $\{
Q_{\bf X}{_i} \}$, where $\mX_i$, $i = 1 \ldots n$, are the set of
7-branes. From now on we shall refer to an equivalence class of
junctions simply as a junction, and denote it as $\mJ$.  Thus the
space of junctions is a lattice of dimension equal to the number of
branes, with a junction $\mJ$ expressed as
\begin{eqnarray}
\label{b}
\mJ = \sum_{i=1}^n \, Q_{{\bf X}_i} \, \mx_i \equiv Q^\mu \,
\mx_\mu\,,
\end{eqnarray}
where $\mu$ indexes the branes and the ``basis strings'' $\mx_\mu$ can
be thought of as a basis of string prongs leaving each brane.

A given junction will carry some total amount of $p$ and $q$ charges
away from the 7-brane configuration; we call these the asymptotic
charges of the junction. The integer $\ell$ introduced in \myref{l}
constrains the possible asymptotic charges that can be realized on
junctions in a given 7-brane configuration. Let ${\bf z}=\ch{p}{q}$
denote the asymptotic charges of a junction emerging from a
configuration, and let
${\bf z}_{i}=\ch{p_i}{q_i}$ denote the brane charges in the configuration.

The constraint takes the form
\begin{eqnarray}
{\bf z} \times \, {\bf z}_{i} = 0  \, (\hbox{mod}\,  \ell )\,\,,
\label{lconstrain}
\end{eqnarray}
for every ${\bf z}_i$ in the brane configuration. Indeed, any
emerging junction  must have ${\bf z} = \sum_i n_i {\bf
z}_{i}$ for integers $n_i$, and then \myref{lconstrain} follows
immediately
from the definition \myref{l}.
In fact, one need only check \myref{lconstrain} for a single brane
${\bf z}_{i}$ of the configuration; if
${\bf z} \times \, {\bf z}_{i}$
and
${\bf z}_{i} \times \, {\bf z}_{j}$ vanish mod$(\ell)$, then
${\bf z} \times \, {\bf z}_{j}$ will also vanish mod$(\ell)$.

Instead of using invariant charges as in \myref{b}, we can specify a
junction with a Lie algebra weight vector $\lambda = a_i \, \omega^i$,
and asymptotic charges $p$, $q$:
\begin{eqnarray}
\mJ = \sum_i a_i\, \momega^i+ p\, \momega ^p + q \, \momega^q \,,
\label{Jinbasis}
\end{eqnarray}
where the $\momega^i$ are $n-2$ junctions of zero asymptotic charge
representing the weights of the algebra, and $\momega^p$ and
$\momega^q$ are Lie algebra singlets with asymptotic charges $(1,0)$
and $(0,1)$ respectively.  The $a_i$ and $p,q$ are linear combinations
of the $Q_{\bf X}{_i}$.

In a few cases, the Lie algebra is not semisimple, but consists of a
semisimple part of rank $r$ and a number of $u(1)$ factors.  In this
case the $(n-2)$ Dynkin labels are replaced by $r$ Dynkin labels and
$((n-2) - r)$ $u(1)$ charges in a generalized weight vector.  For
simplicity of notation, we denote all these charges by $a_i$.  The
asymptotic charges $p$ and $q$, however, are not considered $u(1)$
charges.  The $\momega^i$ associated to $u(1)$ charges will not be
true Lie algebraic fundamental weights, but will just be some basis
junctions with zero asymptotic charge.  Similarly, their duals
$\malpha_i$ will not be true simple roots, but will obey $(\malpha_i,
\malpha_i) <2$; as we shall see shortly, this means they are not
BPS, but they are still useful as basis junctions.

Furthermore, note that $\momega^i, \momega^p, \momega^q$ are
``improper'' junctions, meaning they have non-integral $Q_{\bf
X}{_i}$; a physical junction must combine them in such a way as to be
proper, which places constraints on the asymptotic charges depending
on the conjugacy class of $\lambda$, as discussed in
\cite{dewolfezwiebach}.  Junctions with zero asymptotic charge begin
and end on the 7-branes, while those with nonzero $(p,q)$ carry charge
away from the configuration, perhaps to a probe 3-brane or some other
set of 7-branes.  The junctions with zero asymptotic charge represent
the root vectors of the Lie algebra.

In the M-theory picture of the 7-brane setup, BPS junctions are viewed
as M2-branes wrapped on holomorphic 2-cycles of an elliptically
fibered K3.  A junction supported only on the 7-branes has zero
asymptotic charge; it corresponds to a 2-cycle without a boundary and
defines an element of the second homology group of the K3.  The BPS
representative of an equivalence class is the holomorphic cycle in
that homology class. The lattice property of junctions is consistent
with the abelian nature of the homology group.  As a result, the
lattice of junctions inherits an inner product $(\mJ, \mJ')$ from the
intersection of the corresponding 2-cycles.

For an arbitrary junction $\mJ$ on a 7-brane configuration with
associated finite algebra $\gg$, characterized by a weight vector
$\lambda$ and asymptotic charges $(p,q)$, the self-intersection is
\cite{dewolfezwiebach}:
\begin{equation}
({\bf J},{\bf J}) = - \ll + f(p,q) \,.
\label{intersection}
\end{equation}

The first term is an inner product on the weight vector, $\ll = a_i \,
A^{ij} \, a_j$, where $A^{ij} = -(\momega^i,\momega^j)$.  When the
associated algebra is semisimple, $A^{ij}$ is the inverse Cartan
matrix, since the $\momega^i$ are chosen to be dual to the simple
roots $\malpha_i$, and thus $\ll$ is just the usual Lie algebra inner
product.  When the algebra contains $u(1)$ factors,
$A^{ij}$ is not an inverse Cartan matrix, and
its elements must be determined explicitly by the intersection of
junctions.

In the second term, $f(p,q)$ is a
binary quadratic form in the asymptotic charges given by
\begin{equation}
\label{fpqiu}
f(p,q) \equiv  p^2\,({\mathbold \omega}^p,{\mathbold \omega}^p)+
2pq\, ({\mathbold \omega}^p,{\mathbold \omega}^q)+
q^2\, ({\mathbold \omega}^q,{\mathbold \omega}^q) \,.
\end{equation}
The singlets $\momega^p$ and $\momega^q$ can be represented as a loop
around the 7-branes with an asymptotic string, and $f(p,q)$ can be
derived completely from the monodromy, as we explain shortly.

Notice that although (\ref{intersection}) is always an integer, the
two terms $-\ll$ and $f(p,q)$ individually need not be.  The
requirement that they sum to an integer is another expression of the
conjugacy restrictions on junctions, which restrict the possible
asymptotic charges depending on the conjugacy class of the weight
vector $\lambda$.  The fact that the quadratic form $f(p,q)$, which is
determined entirely from the $\sl2z$ monodromy matrix $K$, has
precisely the form to cancel the non-integral part of the Lie algebra
length-squared $\ll$ is a clue to the interesting connection between
$\sl2z$ conjugacy classes and semisimple Lie algebras.

\subsection{The algebra of junctions}
\label{algebra}

In the Chan-Paton construction of gauge interactions one assigns to
every open string ${\cal S}$ a generator and identifies the structure
constants of the Lie algebra as $f^{{\cal S}_1{\cal
S}_2}_{\;\;\;\;\;\;\;\;{\cal S}_3}=\pm 1$ when the strings
${\cal S}_1$ and ${\cal S}_2$ can be joined to form ${\cal S}_3$. We
now discuss how to generalize this picture when we have string
junctions, and how algebras arise on 7-branes. We will begin with the
generic situation, where we are only able to make general comments,
and then specialize to more familiar cases, where the connection
of junctions with Lie algebras will be more explicit.

The key condition on BPS junctions is that they correspond to homology
cycles that have holomorphic representatives. If the junctions stretch
between 7-branes and thus have no boundaries, a holomorphic
representative exists whenever \cite{wolfson,vsb}
\begin{equation}
\mJ^2 = (\mJ,\mJ) \ge -2\,.
\label{selfad}
\end{equation}
For any {\it nonzero} proper junction (homology class) we introduce a
root space $\hbox{g}_J$. Whenever we deal with finite algebras the
number of roots in each root space is one and we introduce a generator
$E_J$ associated to the root space. In general algebras, however,
there can be a number of different roots in a given root space,
corresponding to a nontrivial root multiplicity, and we must then
introduce generators $E_J^i$, with $i = 1, \cdots m(J)$ for each root
space. The number $m(J)$ is the dimension of $\hbox{g}_J$. The various
roots arise from the same homology class and therefore they are
indistinguishable as junctions. Note that for a nonzero $\mJ$
satisfying (\ref{selfad}), the junction $(-\mJ)$ is also a solution,
so the root spaces $\hbox{g}_J$ and $\hbox{g}_{-J}$ come in pairs.
Moreover, associated to the zero junction, we have a set of Cartan
generators spanning a space $h$.

Let us now consider combining junctions. Given two junctions $\mJ_1$
and $\mJ_2$ with $\mJ_1 + \mJ_2 \not= 0$ and $\mJ_1 + \mJ_2$
satisfying \myref{selfad}, the sum junction can also be realized as a
holomorphic surface and has an associated root space.  We then expect
the Lie algebra bracket to relate root spaces in the usual
way. Therefore, for $\mJ_1 + \mJ_2 \not= 0$ we get
\begin{eqnarray}
\label{liebr1}
[\hbox{g}_{J_1}, \hbox{g}_{J_2}] &\subseteq & \hbox{g}_{J_1+J_2}
\,,\;\;\;\; \hbox{when} \;\;\;\;
(\mJ_1+\mJ_2)^2 \ge -2\, \nn \\ \
[\hbox{g}_{J_1},\hbox{g}_{J_2}] &=& 0 \,,\;\;\;\;\quad\quad \mbox{when}
\;\;\;\;
(\mJ_1+\mJ_2)^2 < -2.
\end{eqnarray}
In addition, we also have
\begin{equation}
[\hbox{g}_{J}, \hbox{g}_{-J}] \subseteq \, h  \,.
\end{equation}
Without further information one cannot list the generators, nor give
structure constants and verify Jacobi identities.

For many (but not all) of the algebras appearing on 7-branes one can
find on the lattice of junctions with zero asymptotic charges a basis
of simple root junctions, $\{ \malpha_i\}, i=1 \ldots r$. These
satisfy $(\malpha_i,\malpha_i) = -2$ (no sum), and have the property
that their intersection matrix defines (minus) a generalized Cartan
matrix (see \cite{kac}).  The Chevalley-Serre construction reproduces
an algebra $\gg$ entirely from its Cartan matrix, and there is an
exact parallel in the framework of junctions.

If $n>0$ copies of the simple root junction $\malpha_i$ can be added
to the simple root junction $\malpha_j$ to obtain a junction that can
be realized holomorphically, we must have:
\begin{equation}
(n\, \malpha_i+\malpha_j ,n \, \malpha_i+\malpha_j) =
-2n^2+2n(\malpha_i,\malpha_j)-2 \ge -2.
\end{equation}
This equation yields
\begin{equation}
n\leq (\malpha_i,\malpha_j) = -A_{ij}\,,
\end{equation}
and it therefore follows that
\begin{equation}
n = 1 - A_{ij} \geq 1 \,,
\end{equation}
is the lowest value of $n$ for which the resulting junction $n\,
\malpha_i+\malpha_j$ cannot be BPS, and therefore is not to generate a
root space. On the Lie algebra side we have the corresponding Serre
relation
\begin{equation}
(\mbox{ad} \, E_{\alpha_i})^{1- A_{ij}}\, E_{\alpha_j} = 0\,,
\end{equation}
which states that $(1-A_{ij}) \alpha_i + \alpha_j$ is not a root.  In
the Chevalley-Serre construction a Lie algebra is completely specified
by the commutation relations of the simple root generators $E_{\pm
\alpha_i}$ and the Cartan generators, together with the Serre
relations. This shows that up to the identification of zero junctions
for Cartan generators, as well as the explicit calculation of root
multiplicities, we have realized the algebra $\gg$ by generators
associated to zero asymptotic charge junctions on the brane
configuration realizing the Cartan matrix of $\gg$ in its intersection
form.  A junction with asymptotic charge will fall into a
representation of $\gg$ and so will be characterized by an appropriate
weight vector $\lambda$.

Finite Lie algebras have all root multiplicities equal to one.  In
fact for all configurations realizing $ADE$ algebras, BPS
junctions with zero asymptotic charge satisfy $\mJ^2= -2$. For each
junction there is a single root, and thus a single root generator
$E_J$.  We then have
\begin{eqnarray}
\label{liebr}
[E_{J_1}, E_{J_2}] & = & \pm E_{J_1+J_2} \,\;\;\;\; \hbox{when}
\;\;\;\;
(\mJ_1+\mJ_2)^2 = -2\,, \nn \\ \,
[E_{J_1},E_{J_2}] &=& 0 \,\;\;\;\;\,\,\,\,\,\,\qquad \mbox{when}
\;\;\;\;
(\mJ_1+\mJ_2)^2 < -2 \,, \\ \,
[E_{J},E_{-J}] &\subseteq & h \,. \nn
\end{eqnarray}
In case of mutually local branes all junctions are necessarily open
strings and \myref{liebr} reproduces the usual Chan-Paton interaction.
In the general $ADE$ case the junctions represent the generators of
$\gg$ in the Chevalley basis, in which all the structure constants are
$\pm 1$. This completes our discussion of the identification of
junctions with Lie algebra generators.

\subsection{The monodromy and the asymptotic charge form $f(p,q)$}
\label{binary}

We now show how the monodromy determines the asymptotic charge form
$f(p,q)$ uniquely.  We shall see in sections 5 and 6 how this
quadratic form not only determines the contribution of a junction's
$(p,q)$ charges to the intersection inner product, but also controls
the enhancement of the algebra as the number of branes is increased.
Furthermore, in section 7 we will see how it plays a role in the
classification of $\sl2z$ conjugacy classes.

Consider a junction $\mJ$ with asymptotic charge ${\bf z}=\ch{p}{q}$,
associated to some 7-brane configuration with monodromy $K$ and Lie
algebra $\gg$.  Let $\mJ$ be a singlet of $\gg$, so $\lambda=0$.  It
can be realized as a string $\overline{\mz}$, crossing the branch cut
to become a $K \overline{\mz}$ string, then joining itself to become
an asymptotic $\mz$-string, where $\mz = K \overline{\mz} -
\overline{\mz}$, as in \figref{loop}.
\onefigure{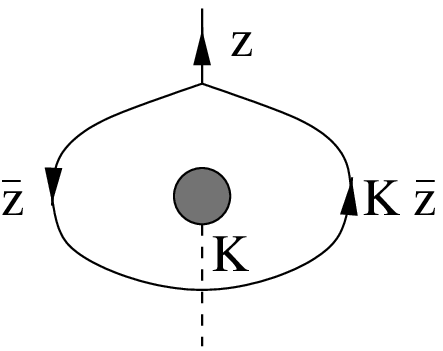}{A singlet
junction as a loop with an asymptotic string of charge $\mz$.}

Every junction that does not intersect the simple roots can be
represented like this, since the roots begin and end on the branes and
so lie within the loop, never crossing it.  Since the $a_i$ vanish,
$\mJ$ is a linear combination of $\momega^p$ and $\momega^q$, as in
(\ref{Jinbasis}).  Rules for computing the self-intersection of
general junctions were discussed in \cite{dewolfezwiebach}.  In this
case, the only contribution comes from the point where the asymptotic
string joins the loop, and is given in terms of the charges of the
string segments by $(\mJ, \mJ) = \mz \mtimes \overline{\mz}$, or
explicitly
\be (\mJ,\mJ) = \mz \times \overline{\mz} = -\mz^T S(K-\unit)^{-1} \mz
\,,
\label{JJquad}
\end{equation}
which by (\ref{intersection})
will be the charge quadratic form
$f_K(\zz)$. Defining $t \equiv {\Tr} K$ and
making use of $\hbox{det}\,(K-\unit)=2- t$ and $t{\unit} = K + K^{-1}$,
we find:
\begin{equation}
f_K({\bf z}) \;\;=\;\; \frac{1}{2-t}\,\,{\bf z}^T S K \mz
\;\;\;\equiv\;\; \frac{1}{2-t}\,\, Q_K(p,q).
\label{ffquad}
\end{equation}
Another useful expression for the charge quadratic form which can be
derived from (\ref{Kdef}) is
\begin{equation}
f_K({\bf z})\;\;=1+\frac{1}{2-t}\{Tr(K_{{\bf z}}K)-2\}.
\ee
This result indicates that the charge quadratic form $f_K(\zz)$ is
determined completely by the $\sl2z$ monodromy matrix of the brane
configuration. More explicitly we can write
\begin{equation}
K=\mat{a}{b}{c}{d} \;\;\;\to
\;\;\; Q_K(p,q)= - c\, p^2+(a-d)\, p\, q + b\, q^2 \,.
\label{matform}
\end{equation}
This map from $\sl2z$ to binary quadratic forms is well known in
mathematics (see for example \cite{Traina}).  Let us take a moment to
explore its properties, as it will turn out to be very useful.  The
map associates $\sl2z$ matrices of trace $t$ to quadratic forms of
discriminant $t^2-4$, and turns out to be one-to-one and invertible.
The inverse map associates to the quadratic form~$Q(p,q)=A \, p^2+B\,
pq+C \, q^2$ the $\sl2z$ matrix of trace $t$ \cite{Traina}
\begin{equation}
K(Q)=\pmatrix{\frac{t+B}{2}& -C\cr A &\frac{t-B}{2}} \,,
\end{equation}
with discriminant $B^2-4AC=t^{2}-4$, and since
$t + B\equiv 0~(\mbox{mod}~ 2)$ the
entries are integral. This map is natural since the fixed points $x$
of $K$ acting on the upper half plane coincide with the zeroes of
$Q_K(x,1)$:
\begin{equation}
\frac{ax+b}{cx+d}=x \;\;\; \Longleftrightarrow
\;\;\; -cx^2+(a-d)x+b = 0 \,.
\end{equation}
Moreover, the relation \myref{matform} establishes a one-to-one
correspondence between $\sl2z$ conjugacy classes of trace $t$ and
equivalence classes of quadratic forms of discriminant $t^2-4$. In
other words two $\sl2z$ matrices of trace $t$ are conjugate in $\sl2z$
if and only if the associated quadratic forms are equivalent ({\it
i.e.} $Q \sim Q'$ if $Q'(\mz) = Q(g \mz)$ for some $g\in
\sl2z$). Indeed, we have
\begin{equation}
\label{classes}
K' = gKg^{-1} \;\;\; \Leftrightarrow \;\;\;
Q_{K'}({\bf z})=Q_{K}(g^{-1}{\bf z})\,..
\end{equation}
Let us prove this. By explicit substitution the direction
$\Rightarrow$ is straightforward. To see the opposite, consider $K$
and $K'$ satisfying $Q_{K'}(\mz)=Q_{K}(g^{-1} \mz)$ for any $\mz$:
\begin{equation}
\mz^T g^{T-1} S K g^{-1} \mz = \mz^T S K \mz,
\end{equation}
implying
\begin{equation}
SgKg^{-1} = SK+mS,
\label{step2}
\end{equation}
for some $m$, where we used $Sg=(g^{T})^{-1}S$. Multiplying
\myref{step2} by $S$ and taking the trace of both sides yields $m=0$,
proving (\ref{classes}).  This should not surprise us, since our
construction of $f_K(p,q)$ was manifestly $\sl2z$-covariant.

We have shown that the monodromy $K$ entirely
determines the asymptotic charge form $f(p,q)$, and additionally one
can show that from $f(p,q)$ one can uniquely recover $K$.  We see that
the problem of enumerating the conjugacy classes of $\sl2z$, which in
turn determine the possible algebras realized on 7-branes, is
equivalent to that of classifying inequivalent quadratic forms.  We
shall come back to this point later.

\medskip

\subsection{The monodromy and the determinant of $A(\gg)$}
\label{monodet}

In this section, we shall express the determinant of the Cartan matrix
of the semisimple algebra $\gg$ arising on a set of 7-branes in terms
of  two
simple invariants of the 7-brane system, the trace of the monodromy
matrix $K$ and the asymptotic charge invariant $\ell$.  

For any brane configuration $\mG = \mX_1 \mX_2 \ldots \mX_n$,
the metric on the associated 
lattice of junctions $\Lambda$  
is given by \cite{dewolfezwiebach}
\begin{equation}
\A \equiv
\left(\begin{array}{cccc}
1 & a_{12} & \ldots &a_{1n} \\
a_{12} & 1 & \ldots &a_{2n} \\
\vdots & \vdots & \ddots & \vdots \\
a_{1n} & a_{2n}& \ldots &1
\end{array}\right) \hspace{.5in}
a_{ij} = -\frac{1}{2}(\mathbold{z}_i\times \mathbold{z}_j)=
-\half(p_iq_j-q_ip_j).
\label{basisdeff}
\end{equation}
The volume of the unit cell in the total junction lattice $\Lambda$ is
given by
$\sqrt{\mbox{det}\A}$, and one finds that this quantity only depends on
the trace of
the overall monodromy:
\begin{eqnarray}
\mbox{det}\,\A = \frac{1}{4}\,\,\hbox{Tr}\, K +\frac{1}{2}\,\,,
\label{trdet1}
\end{eqnarray}
This equation, which holds for any brane configuration, is proven using
\myref{threemon} and
\myref{basisdeff}. Since our proof is somewhat technical we have
relegated it to the
appendix.

The lattice $\Lambda$ has a sublattice $\Lambda_0$ containing all
junctions with no asymptotic charges. We denote by $\A_0$ the matrix
giving the metric on $\Lambda_0$ for some specific choice of basis.
Being the volume of the unit cell, $\sqrt{\mbox{det}\A_0}$ is basis
independent. It is shown in the appendix that 
\begin{eqnarray}
\ell^2\;\mbox{det}\A_0 = 2- {\Tr} K \,.  
\label{tracedett}
\end{eqnarray}

When we have a set of all mutually local branes -- for example
$\mA^{N+1}$ , which realizes $\gg = A_N$ -- the lattice of asymptotic
charges is only one-dimensional, $\ell=0$ and ${\Tr} K = 2$, in which
case \myref{tracedett} is trivially satisfied. This situation is
treated separately in the appendix where we find
\begin{eqnarray}
\mbox{det}\A_0 = N+1 \,, \qquad
\hbox{for}\,\,(N+1) \,\,\hbox{mutually local branes} \,.
\label{maineqA}
\end{eqnarray}


We now recall that the Lie algebra $\gg$ associated to a brane
configuration \mG\ is precisely identified by matching the basis basis
junctions of $\Lambda_0$ to the simple roots of $\gg$ in such a way
that inner products coincide \cite{dewolfezwiebach}.  This is exactly
the situation when $\gg$ is semisimple, in which case simple roots
describe the algebra completely. Then $\mbox{det}\A_0$ coincides with
$\mbox{det}A(\gg)$, the determinant of the Cartan matrix of
$\gg$. Indeed, \myref{maineqA} is consistent since the $A_N$ Cartan
matrix has determinant $N+1$.  For \myref{tracedett} we write
\begin{eqnarray}
\ell^2\;\mbox{det}A(\gg) = 2- {\Tr} K \,.
\label{tracedet}
\label{maineq2}
\end{eqnarray}
Here, however, we must note that when $\gg$ is not semisimple
$A(\gg)$ is not a Cartan matrix. Consider, for example,
the case $\gg= \bar\gg \oplus u(1)$, with
$\bar\gg$ semisimple. Assume basis junctions $(\{ \mJ_i\}, \mJ)$
can be found such that the $\{\mJ_i\}$ generate $\bar\gg$,
$\mJ$ carries the $u(1)$ charge, and $\mJ \cdot \mJ_i =0$. Then
$\mbox{det} A(\gg) = \mbox{det}A (\bar\gg) \cdot (-\mJ^2)$.


We will use the trace/determinant relation (\ref{tracedet}) in section
\ref{classify} to fit configurations with various algebras into
conjugacy classes of $\sl2z$.  For now, let us consider what it tells
us about the algebras realized on the configurations ${\bf E_1}$ and
${\bf \tilde{E}_1}$, which as we mentioned have equivalent $K$ but
different $\ell$.

We determined in section \ref{monodromy} that $\ell({\bf E_1}) = 2$.
Its monodromy is $K({\bf E_1}) = \left( -2 \; -7\atop -1 \;
-4\right)$, and consequently, ${\Tr} K = -6$.  Using (\ref{tracedet}),
we learn that det$A(\gg({\bf E_1})) = 2$.  
The only $ADE$ algebra with such a
Cartan matrix is $\gg({\bf E_1}) = A_1$.  Indeed, explicit examination
of the junctions supported on ${\bf E_1}$ reveals a single simple root
with $\mJ^2 = -2$.

The monodromy of ${\bf \tilde{E}_1}$ is conjugate to that of ${\bf
E_1}$, and consequently ${\Tr} K({\bf \tilde{E}_1}) = -6$ as well;
however, $\ell({\bf \tilde{E}_1}) =1$.  Equation (\ref{tracedet}) then
tells us that det$A(\gg ({\bf \tilde{E}_1})) = 8$.  Thus the algebra
cannot
be $A_1$.  However, the configuration has only three branes, and
cannot support an algebra with rank greater than one.  The only
possible conclusion is that $A(\gg ({\bf \tilde{E}_1}) )$ is not the
Cartan
matrix of a semisimple Lie algebra, but instead is just the
intersection form of a basis vector corresponding to an Abelian
factor.  Indeed, the minimal uncharged proper junction on the
configuration has self-intersection $\mJ^2 = -8$, and so $\gg({\bf
\tilde{E}_1}) = u(1)$.

The 7-brane configurations corresponding to $E_N$ can be used to
construct ${\cal N}=2$ \mbox{$D=4$} theories with exceptional global
symmetries which are $S^1$ compactifications of the five dimensional
theories with the same global symmetry
\cite{seibergfive,MS,AHK,leungvafa}.  Therefore the presence of two
configurations either of which can enhance to $E_2$ is consistent with
the fact that in five dimensions there are two different theories,
$E_1$ and $\tilde{E}_1$, with $su(2)$ and $u(1)$ global
symmetry. These two theories become equivalent to the $E_2$ theory
after addition of more matter. In \cite{MS} these theories were
related to shrinking del Pezzo surfaces in a Calabi-Yau threefold. In
that framework the two theories $E_1$ and $\tilde{E}_1$ correspond to
the two del Pezzos $\bbbp^{1} \times \bbbp^{1}$ and ${\cal B}_1$,
where ${\cal B}_n$ is $\bbbp^{2}$ blown up at $n$ generic points. It
is a known fact that further blowing up either one at a point gives
the manifold ${\cal B}_2$.  Correspondingly, the ${\bf E_1}$ and ${\bf
{\tilde{E}_1}}$ configurations become equivalent as ${\bf E_2}$ after
addition of a D7-brane.

\section{From Kodaira Singularities to Infinite Series}
\label{kodaira}

The Kodaira classification of singularities on a K3 manifold
tells us that in certain limit of moduli space, a collection
of 2-cycles with intersection realizing an $ADE$ Cartan matrix
collapse to zero size.  In the 7-brane picture, this means
that certain sets of 7-branes can be brought to a point.  Other
configurations, with other possible algebras realized by the
intersection form, may not, but are still interesting to study.

In this section we generalize the configurations corresponding to
Kodaira singularities into a number of infinite series, recognizing a
more systematic way of treating these series in the process.
We begin the discussion by examining configurations of two 7-branes,
which turn out to be ``kernels'' for the infinite series we discuss
in the second half of the section.

\subsection{Classification of configurations of two 7-branes}
\label{twobranes}

The complete enumeration of inequivalent configurations of two
7-branes is still a difficult problem, but we shall organize the
classification, and examine the first few nontrivial cases.

A pair of 7-branes may be either mutually local or nonlocal.  Two
mutually local branes are always $\sl2z$-equivalent to $\mA_1 = \mA
\mA$, which realizes the $A_1$ algebra and supports only $p$ charge.

Two mutually non-local branes will support junctions with both $p$ and
$q$ charge.  There are no junctions without asymptotic charge and no
enhanced symmetry algebra.  If the charges of the 7-branes are
$[p,q]$, and $[r,s]$, then $\ell = ps-rq$ from (\ref{l}).  Furthermore
the trace of the monodromy is readily computed to be
\begin{equation}
{\Tr} K = 2 - \ell^2 \,.
\label{2trace}
\end{equation}
Note that in larger brane configurations ${\Tr} K$ need not depend on
$\ell$ in any fashion; indeed ${\bf E_1}$ and ${\bf \tilde{E}_1}$ have
identical $K$ despite differing $\ell$.  The relationship
(\ref{2trace}) is special to the case of two 7-branes.

We have conjectured that only the equivalence class of $K$, the value
of $\ell$ and the number of branes will classify any configuration of
7-branes.  In this instance we have fixed the number of branes, and
since $\ell$ follows from $K$, we predict that only the equivalence
classes of the monodromy will specify equivalence.

Two configurations will only be equivalent if $\ell$ coincides.  By
separate global $\sl2z$ transformations we can always convert them
into the form $\mA \mX_{\bf [i,\ell]}$, $\mA \mX_{\bf [j,\ell]}$
where $0\leq i,j\leq~\ell$.  Configurations with the same $\ell$ may
be equivalent if the values $i$, $j$ can be mapped into each other
by transformations preserving the form of these configurations.

Since the integers $(i, \ell)$ characterize a 7-brane, we must have
gcd$(i,\ell)$=1. In this case there exist integers $(a,b)$ such that
$a i -b \ell = 1$, using which we can construct an $\sl2z$ matrix:
\begin{equation}
g=\pmatrix{a&-b\cr -\ell&i}.
\end{equation}
We then apply this transformation, move a branch cut and
apply another $\sl2z$ by some power of $K_A = \left({1\; -1 \atop
\!0\;\;\; 1}\right)$:
\begin{equation}
\label{mstusee}
\mA\mX_{\bf [i,\ell]}\stackrel{g}{\cong} \mX_{\bf [-a,\ell]}\mA\, =
\mA\mX_{\bf [-a-\ell,\ell]}
\stackrel{(K_A)^n}{\cong} \mA\mX_{\bf [a^*,\ell]} \,,
\end{equation}
where in the power $n$ of $K_A$ is chosen to as to obtain
\begin{equation}
\label{mstuse}
0\leq a^*\leq \ell\,, \qquad \hbox{with} \quad a^* =
-a \quad (\hbox{mod} \, \ell)\,.
\end{equation}
This result helps the classification as follows. For a given $\ell$
the above relation will provide equivalences between $\mA \mX_{\bf [i,\ell]}$ 
configurations with different allowed values of
$i$. Configurations that cannot be made equivalent in this way, may
or may not be equivalent.  Let us now catalog the first few cases.

The value ${\Tr} K = +2$ corresponds the two 7-branes being are
mutually local. The brane configuration is then always equivalent to
an ${\bf AA}$ configuration and the resulting gauge algebra is
$A_1$. As an aside, since affine algebras can only arise for ${\Tr}
K=+2$ (see \cite{dewolfe,infinite}) this implies that affine algebras
cannot arise with just two 7-branes.

The case $\ell=1$ gives ${\Tr} K=1$ and can be achieved with $\mA
\mX_{\bf [i,{\bf 1}]}$. Here the two possible cases $i=0$ and $i=1$ are
manifestly equivalent by suitable $K_A$ conjugation.  We find it
convenient to choose the representative ${\bf H_0} = \mX_{\bf [0,
-1]}{\bf C}$, the first element of the Argyres-Douglas series, and a
Kodaira configuration in its own right.  We can show it is equivalent
to the canonical form by $\mX_{\bf [0, -1]}{\bf C} = \mA \mX_{\bf
[0,-1]} = \mA \mX_{\bf [0,1]}$.

For $\ell=2$, ${\Tr} K=-2$ and the unique realization is $\mA \mX_{\bf
[1,2]}$ since $i=0,2$ do not denote good 7-branes. We can represent
this case by the ${\bf D_0} = {\bf BC}$ system, as global action of
$g=\left({1\; 0 \atop 1\; 1}\right)$ on ${\bf BC}$ gives us
$\mA\mX_{\bf [1,2]}$. ${\bf D_0}$ is the background 7-brane
configuration for realizing the familiar $SU(2)$ Seiberg-Witten theory
with $N_f = 0$ on the world volume of a D3-brane probe.

The case $\ell=3$ giving ${\Tr} K=-7$ can be realized as $\mA\mX_{\bf
[1,3]}$ and as $\mA\mX_{\bf [2,3]}$.  It is simple to verify using
(\ref{mstusee}), (\ref{mstuse}) that these two are actually
equivalent. We can therefore choose to represent this case by ${\bf
E_0} = {\bf X_{\bf [2,-1]} C}$, a configuration with $\ell=3$.
This will be the kernel for the exceptional series.

We do not find two inequivalent configurations until $\ell=5$, where
${\Tr} K=-23$.  The brane configurations are $\mA\mX_{\bf
[1,5]}\cong\mA\mX_{\bf [4,5]}$ and $\mA\mX_{\bf [2,5]}\cong\mA\mX_{\bf
[3,5]}$. Inequivalence is proved explicitly by a calculation showing
that the monodromies are not conjugate in $\sl2z$.

One can continue in this fashion without encountering much new.  Let
us now turn to the infinite series we can generate from these two
7-brane kernels.

\subsection{Infinite Series}

The examples of $A_N$, $D_N$ and $E_N$ Kodaira singularities
\cite{Kodaira} have been studied considerably, and have been
associated with the 7-brane configurations \cite{dasguptamukhi/sen}:
\begin{eqnarray}
{\bf A}_{\bf N} &:& {\bf AA\ldots A} \equiv {\bf A}^{N+1} \nn\\
{\bf D}_{\bf N\ge4} &:&  {\bf A}^N{\bf BC} \\
{\bf E}_{\bf 6},{\bf E}_{\bf 7},{\bf E}_{\bf 8} &:& {\bf A}^5{\bf
BCC}, {\bf A}^6{\bf BCC}, {\bf A}^7{\bf BCC} \nn \\
{\bf H}_{\bf 0},{\bf H}_{\bf 1},{\bf H}_{\bf 2} &:&
{\bf AC}, {\bf AAC}, {\bf AAAC}.  \nn
\end{eqnarray}
The ${\bf D}_{\bf 4}, {\bf E}_{\bf 6},{\bf E}_{\bf 7},{\bf E}_{\bf 8}$
correspond to orbifold singularities of the K3, and the
${\bf H}_{\bf 0},{\bf H}_{\bf 1},{\bf H}_{\bf 2}$ are
the Argyres-Douglas points with associated algebras ${\cal G}
=0,A_1,A_2$.  These configurations are ``dual'' to the ${\bf E_N}$
 in the sense that ${\bf E_N}$ and ${\bf H_{8-N}}$ can be realized
with the same constant value of $\tau$ and together the net monodromy
is unity.

These configurations naturally suggest generalizations.  The
non-collapsible ${\bf D_N}$ configurations with $N = 3,2,1,0$ support
algebras $D_3=A_3$, $D_2=A_1\oplus A_1$, $D_1=u(1)$ and $D_0=0$, and
are backgrounds of a probe D3-brane realization for the Seiberg-Witten
theories with $N$~flavors, where the 7-brane algebra is realized
as a global symmetry.

Similarly, we can extend ${\bf E_N}$
to ${\bf A}^N{\bf BCC}$ with $N = 1,2,3,4,5$, realizing the algebras
$E_1=A_1$, $E_2=A_1\oplus u(1)$, $E_3=A_2\oplus A_1$, $E_4=A_4$, and
$E_5=D_5$.  We have already seen that ${\bf E_5} \cong {\bf D_5}$, and
one can also confirm ${\bf E_4} \cong {\bf H_4}$.  Continuing the
${\bf E_N}$ series beyond $N=8$ one encounters ${\bf E_9}$ which gives
rise to the affine algebra $E_9 =\widehat{E}_8$ \cite{dewolfe}, which
we shall have more to say about in the sequel paper \cite{infinite}.

The $\sl2z$ equivalence transformations are also useful in unifying
the description of the ${\bf D}$, ${\bf E}$, ${\bf H}$ series which
gives hint to further generalization. It is straightforward to check
the following identity:
\begin{eqnarray}
{\bf A}^{N+1}{\bf C} \stackrel{K_A}{\cong} {\bf A}^{N+1}
{\bf X_{\bf [0,-1]}} =
{\bf A}^{N}  {\bf X_{\bf [0,-1]} C} \,,
\end{eqnarray}
which allows us alternately to describe this series as
${\bf \tilde{H}_N} = {\bf A}^N  {\bf X_{[0,-1]}  C }\,$.
Thus the kernel ${\bf H_0}$ from the previous section generates
the entire Argyres-Douglas series just by adding $\mA$-branes.

Similarly, the ${\bf D_N}$ series is described by ${\bf D_N}= {\bf
A}^N {\bf X_{[1,-1]} C }\,,$ where we have written explicitly the
charges of the $\mB$-brane.  It is generated from the ${\bf D_0}$
kernel.  Notice that although $\ell({\bf D_0}) = 2$, $\ell({\bf D_N})
= 1$ for $N >0$.

This now suggests a similar possibility for the ${\bf E_N}$ series.
Indeed, we noted in (\ref{enequiv}) that
\begin{eqnarray}
{\bf A}^{N-1}{\bf BCC} &\cong& {\bf A}^N {\bf X_{[2,-1]}  C},
\end{eqnarray}
where the conjugation is with the matrix $K_A$. This enables us to
write another presentation for the $E_N$ series as ${\bf \tilde{E}_N}
= {\bf A}^N {\bf X_{[2,-1]} C }\,$.  We see that the ${\bf E_0}$
kernel does indeed generate the entire series. As we discussed in
section \ref{monodromy}, the ${\bf E_N}$ and ${\bf \tilde{E}_N}$
series are only equivalent for $N\geq 2$, while ${\bf E_1}$ and ${\bf
\tilde{E}_1}$ are not equivalent.

With these examples in mind, we recognize the significance of
the Kodaira series.  ${\bf A_N}$ may be thought of as the series
that results from adding a number of ${\mA}$-branes to a kernel of
a single ${\mA}$-brane.  Similarly, ${\bf H_N}$, ${\bf D_N}$
and ${\bf E_N}$ are the three simplest series generated from kernels
of two branes.

The three series of configurations can be uniformly described as:
\begin{equation}
{\bf S_{p,N}} = {\bf A}^N{\bf X_{[p,-1]} C}\,,
\end{equation}
with $p=0,1,2$ for the ${\bf \tilde{H}_N}, {\bf D_N}$ and ${\bf
\tilde{E}_N}$ series, respectively. Notice that in all cases,
$\ell = 1$ for $N \geq 1$.

To see how the Lie algebras arise, we show in \figref{simple} the
simple root junctions whose intersection form produces the
corresponding Cartan matrix. The trace of the overall monodromy is
\begin{equation}
{\Tr} K({\bf S_{p,N}}) = 2-(p+1)^2+(p-1)N \,,
\end{equation}
so that $2-{\Tr} K(p,N)=(p+1)^2-(p-1)N$ which reproduces the expected
relation to the determinant of the Cartan matrices:
det$\,(A(A_N))=N+1$, det$\,(A(D_N))=4$ and det$\,(A(E_N))=9-N$.
\onefigure{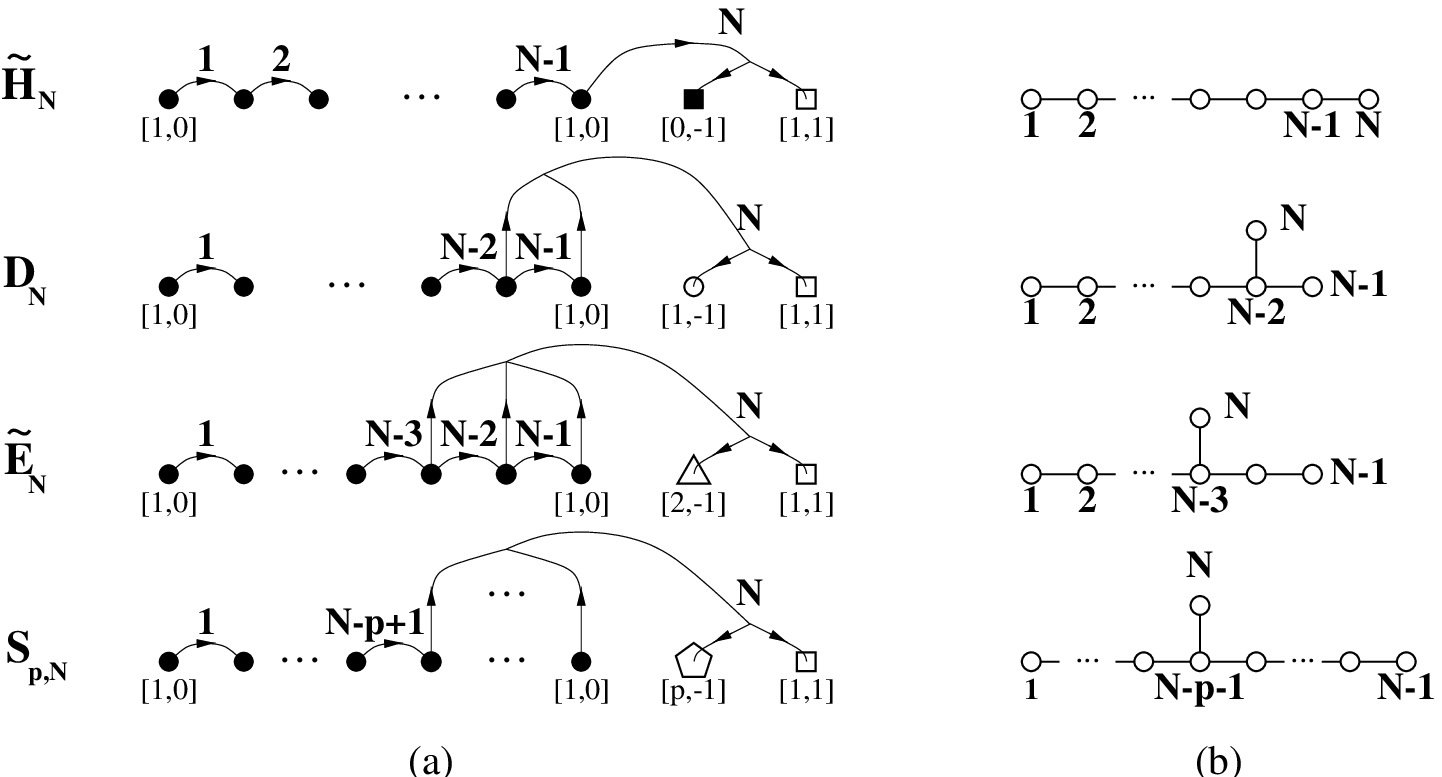}{(a) Brane configuration of the ${\bf H}$,
${\bf D}$, ${\bf E}$ and
the generalized ${\bf S_p}$ series. (b) Dynkin diagrams of the algebras
$A_N$, $D_N$, $E_N$ and $T_{p+1,2,N-1-p}$.}

Notice that besides the series of finite $A$ and $D$ algebras, we have
constructed $E_N$ whose elements are infinite dimensional Kac-Moody
algebras for $N>8$. In fact one could go further by engineering new
algebra realizations with $p>2$, those Dynkin diagrams are shown in
\figref{simple}(b) and are usually denoted as
$T_{p+1,2,n-1-p}$. Finally the simple root junctions suggest how to
the realize $\gg = T_{r_1,r_2,r_3}$: the configuration is ${\bf
A}^{r_1+r_2}{\bf X_{[r_1-1,-1]}}{\bf C}^{r_2-1}$, the simple roots are
shown in \figref{Trrr}(a), and one can check that \myref{tracedet} is
satisfied.  We shall not pursue these exotic series further; no doubt
there is more to be said.  We shall have more to say about
infinite-dimensional algebras in \cite{infinite}.
\onefigure{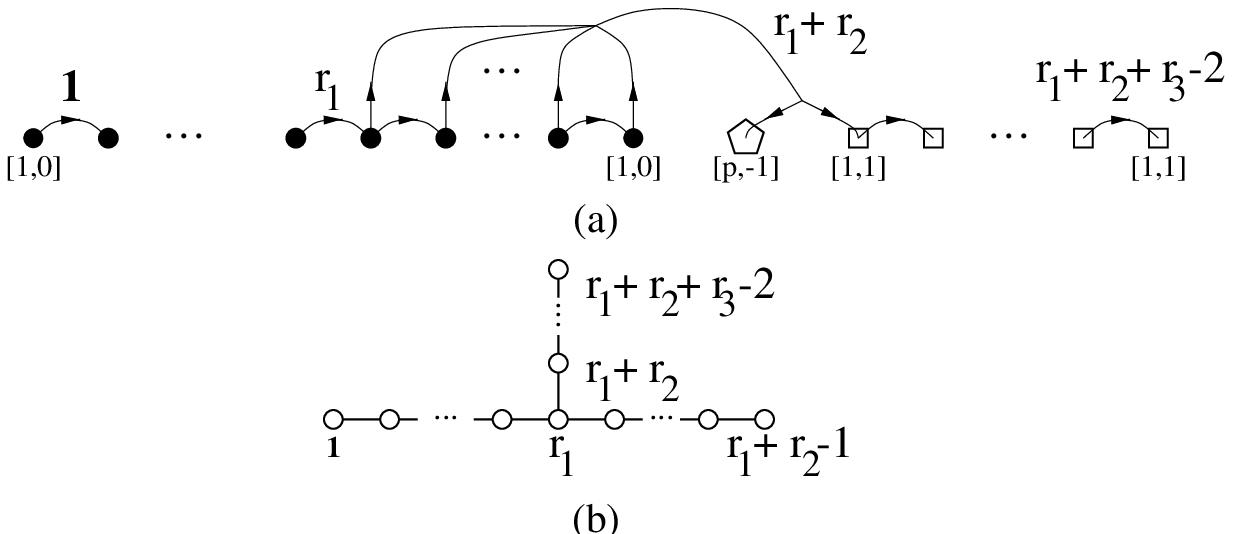}{(a) Brane configuration of $\gg = T_{r_1,r_2,r_3}$.
(b) Dynkin diagram of $\gg = T_{r_1,r_2,r_3}$.}

It is remarkable that we have found two series, ${\bf A_N}$ and ${\bf
H_N}$, realizing the $A_N$ algebras, the former supporting junctions
with only $p$ charge and $\ell=0$, the latter with both $p$
and $q$ asymptotic charges and $\ell = 1$.  In fact there are an
infinite number of such series, parameterized by $\ell$.

Take any kernel of two 7-branes with some value $\ell$, and perform an
$\sl2z$ transformation such that one of the 7-branes is an
$\mA$-brane, ${\bf \mA \mX_{[i,\ell]}}$, as in the previous subsection.
One may add additional $\mA$-branes without modifying $\ell$, to
obtain a series
\begin{eqnarray}
{\bf A^{{\bf i}\,\ell}_N} \equiv \mA^{N+1} \mX_{\bf [i,\ell]} \,, 
\end{eqnarray}
which has $\gg = A_N$ and asymptotic charge invariant $\ell$.
Configurations with different $\ell$ are obviously inequivalent, thus
proving there are an infinite number of series realizing $A_N$.  We
have not found any analogous configurations for the $D_N$ and $E_N$
algebras, which seem to appear on just one series each.

Notice that, like the ${\bf S_{p,N}}$ series, the ${\bf A_N^{{\bf
i}\,\ell}}$ series arise from adding $\mA$-branes to a kernel of two
7-branes with some $\ell$.  They differ in the $\sl2z$ presentation of
that kernel, relative to the $\mA$-branes.

We shall end our discussion of series of 7-branes here, since we have
found more than enough to occupy our attention.  This technique of
beginning with a known configuration and enhancing it by adding a new
brane is a powerful one, and offers insight into how the monodromy
fixes the algebra by means of the quadratic form.  We shall explore it
systematically in the next section.

\begin{table}
\begin{center}
\begin{tabular}{|c|l|c|c|} \hline
 $\mathcal{G}$ & Brane Configuration & $K$ & $f_{K}(p,q)$
\\ \hline \hline
$\rule{0mm}{7mm}$ $A_{N}$
& ${\bf {A}_N}={\bf A}^{N+1}$
& $\pmatrix{1&-N-1\cr 0&1}$
&$-\frac{1}{N+1}p^{2}$   \\ \cline{2-4}

&$\rule{0mm}{7mm}$ ${\bf {H}_N}={\bf A}^{N+1}{\bf C}$
& $\pmatrix{2&-3-2N\cr 1&-N-1}$
&$\frac{1}{N+1}\{-p^{2}+(N+3)pq-(3+2N)q^{2}\}$   \\ \cline{2-4}

& $\rule{0mm}{7mm}{\bf \tilde{H}_N}={\bf A}^{N}{\bf X}_{\bf [0,-1]}{\bf
C}$
& $\pmatrix{1&-N-1\cr 1&-N}$
&$\frac{1}{N+1}\{-p^{2}+(N+1)pq-(N+1)q^{2}\}$   \\ \hline

$\rule{0mm}{7mm}$$D_{N}$  & ${\bf D_N}={\bf A}^{N}{\bf BC}$
& $\pmatrix{-1&N-4\cr0 &-1}$
& $\frac{N-4}{4}q^{2}$  \\ \hline

$\rule{0mm}{7mm}$$E_{N}$  & $\rule{0mm}{7mm}$ ${\bf E_N}=
{\bf A}^{N-1}{\bf BCC}~$
& $\pmatrix{-2 & 2N-9\cr -1 &N-5 }$ &
$\frac{1}{9-N}\{p^{2}+(3-N)pq+(2N-9)q^{2}\}$ \\ \cline{2-4}
   &$\rule{0mm}{7mm}$ ${\bf \tilde{E}_N}={\bf A}^{N}
             {\bf X}_{\bf [2,-1]}{\bf C}$
& $\pmatrix{-3&3N-11 \cr -1 &N-4}$
& $\frac{1}{9-N}\{p^{2}+(1-N)pq+(3N-11)q^{2}\}$ \\ \hline
\end{tabular}
\end{center}
\caption{Brane configurations, monodromies and the charge quadratic
form for $A_{N}, D_{N}$ and $E_{N}$ algebras. The two series (${\bf
H_N}$ and ${\bf \tilde{H}_N}$) realizing the $A_N$ algebras are
equivalent. The two series (${\bf E_N}$ and ${\bf \tilde{E}_N}$)
realizing $E_N$ are equivalent for $N\ge2$.}
\end{table}
\newpage

\section{Transitions Between 7-Brane Configurations}

We have examined several series of finite Lie algebras arising on
7-branes.  Keeping in mind the general question of understanding all
possible configurations, in the present section we address transitions
from one algebra to another.  Here we shall see how the charge
quadratic form $f(p,q)$ controls the change from one algebra to the
other.  Suppose we have a brane configuration $\mG$ with monodromy $K$
and quadratic form $f_K$ that realizes a finite Lie algebra $\gg$.
When we add one more 7-brane $\mZ$ with charge $\mz = [p,q]$, we
obtain a new configuration $\Genh= \mG\mZ$, where we have
conventionally placed the new brane on the right of the configuration
(see \figref{ajunction}). Additional junctions with support on the new
brane appear, resulting in an enhancement from $\gg$ to some larger
algebra $\genh$.

Which algebra $\genh$ is obtained depends on $\gg$ and on the charge
$\mz$ of $\mZ$.  Recall that \mbox{$f(\mz)=f(\mz')$} when $\mz' = g
\mz$ with $g \in \sl2z$ and $g K g^{-1} = K$.  As a result, the total
monodromy $K_{\mz} \, K$ of $\Genh$ is conjugate to $K_{\mz'} \, K$,
and we expect both $\mZ$ and $\mZ'$ to enhance $\gg$ to the same
$\genh$.  The enhancing 7-branes can therefore be grouped into
classes; all elements of a given class give the same enhancement.  The
charge quadratic form $f(\mz)$ thus measures how the monodromy
``sees'' the charges of the enhancing brane.  In general
$\mbox{rank}(\genh) = \mbox{rank}(\gg) + 1$ (except when the initial
configuration contains only mutually local branes, in which case
adding a mutually nonlocal brane will not enhance $\gg$) and so the
enhancing brane opens up a new direction in the weight lattice.

\onefigure{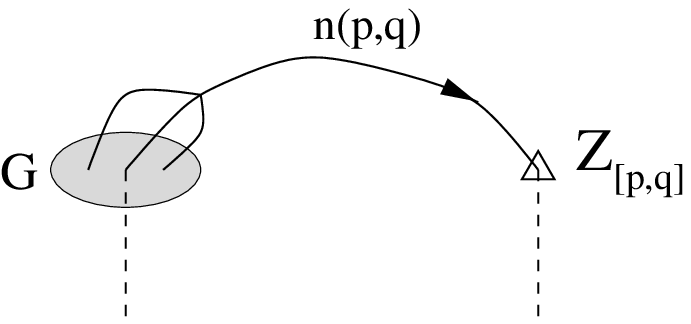}{The enhanced configuration $\Genh$ is obtained
by adding a ${\bf Z_{[p,q]}}$-brane to the original configuration
$\mG$. A generic junction has support on the new brane.}
To investigate the enlarged root lattice, we construct all
junctions on $\Genh$ having zero total asymptotic charge.  Each can be
written
\begin{eqnarray}
\mJ = -n\ \mz + n\,( p\,\momega^p + q\, \momega^q) + \sum_i a_i \,
\momega^i  \,,
\end{eqnarray}
where the $\momega^i$ are the weight junctions for $\gg$ and have zero
asymptotic charge, and $\{ \momega^p, \momega^q \}$ are the singlets
of $\gg$ with asymptotic charges $\mz_p = (1,0)$ and $\mz_q = (0,1)$.
This junction can be visualized as a sub-junction which leaves the
$\mG$ configuration with charges $(np,nq)$ and ends on the $\mZ$
brane, where it has $n$ prongs, as in \figref{ajunction}. The
self-intersection of $\mJ$ is readily calculated, since there are no
cross-terms between the $\mG$ sub-junction and the $\mZ$-brane prongs.
We find
\begin{eqnarray}
\mJ^2 =  - n^2  + n^2 f(\mz) - \ll \,,
\label{pqs}
\end{eqnarray}
where $\mJ^2 = (\mJ, \mJ)$, and $\lambda = \sum_i \, a_i \, \omega^i$
is the $\gg$ weight vector.  A minor rearrangement gives
\begin{eqnarray}
\ll = -\mJ^2 + n^2 (f(\mz) - 1) \,.
\label{ll}
\end{eqnarray}
New junctions appear whenever this equation has a solution with
non-vanishing grade $n$, and we identify them as the roots of the
algebra $\genh$.  For each value of $\ll$ there are a number of weight
vectors $\lambda$, filling out a Weyl orbit of $\gg$, and there will
be a distinct root of $\genh$ for each distinct weight $\lambda$.
Thus each $\gg$ root junction can be characterized by $\malpha = (n,
\lambda)$.  Since $\ll \geq 0$ always and supersymmetry requires
$\mJ^2 \geq -2$, it is possible that (\ref{ll}) cannot be satisfied
for some $n$, in which case there are simply no root junctions at that
$n$.  Naturally the roots of $\gg$ appear as $(0,\alpha_{\gg})$.

Since roots have no overall asymptotic charge, it is natural to
interpret equation (\ref{pqs}) as $\mJ^2 = -\LL$, where $\Lambda$ is a
$\genh$ weight vector.  The length squared of the $\genh$ weight
vector is
given in terms of the length squared of the $\gg$ weight vector plus a
contribution along the new axis in the weight lattice, with $(1 -
f(\mz))$ setting the scale for the new direction.

Once we have determined the root system of $\genh$, we next find a
subset of {\it simple} roots.  This requires ensuring that any root
can be written in the basis of simple roots with coefficients that are
either all positive or all negative integers.  The simple roots then
determine both the Cartan matrix and the Dynkin diagram of $\genh$.
Simple roots of Kac-Moody Lie algebras are always real, meaning the
corresponding junctions have $\mJ^2 = -2$.

Since the rank increases by one, generically we need exactly one new
root in addition to the simple roots of $\gg$ to complete the set of
simple roots for $\genh$.  (A few exotic cases where two (linearly
dependent) new simple roots are necessary are examined in
\cite{infinite};
in all these cases $\genh$ is an infinite-dimensional
algebra.)

When there is a single new simple root $\malpha_0 = (n_0, \lambda_0)$,
it must be that $|n_0|\leq |n|$ for any $n$ with solutions for
(\ref{ll})
since we need to write any root
as an integer linear combination of simple roots, and
all other simple roots have $n=0$.  Let $W_0$ denote
the Weyl orbit at grade $n_0$.  Each root $\malpha = (n_0, \lambda)$
which is associated to a weight $\lambda \in W_0$ must be positive,
since ${\malpha}=\ldots+{\malpha_0}$
(i.e. the coefficient of $\malpha_0$ is already positive).
It is then necessary
that $\lambda - \lambda_0$ be a positive root of $\gg$ for every
$\lambda \in W_0$.  This implies that $\lambda_0$ must be the lowest
weight $\theta_<$ in the Weyl orbit $W_0$.

In the next section we shall explore how the value $f(\mz)$ determines
the type of enhancement that occurs.  We shall consider only
enhancements to finite $\genh$.  Enhancements to infinite-dimensional
$\genh$ are examined in \cite{infinite}.  The structure that we shall
find is summarized in Table~\ref{enhancef}.

\begin{table}
\begin{center}
\begin{tabular}{|c|c|} \hline
$\rule{0mm}{5mm}f(\mz)$ & $\genh$ \\ \hline\hline
 $f(\mz) < -1$ & $\gg \oplus u(1)$ \\
 $f(\mz) =-1$ & $\gg \oplus A_1$ \\
$-1 < f(\mz) < 1$ & finite \\
$f(\mz) =1$ & affine  \\
$f(\mz) >1$ & indefinite \\ \hline
\end{tabular}
\end{center}
\caption{Algebraic enhancements of brane configurations.  Adding one
brane to an existing configuration, the asymptotic charge form $f(\mz)$
of
${\bf G}$ evaluated on the new brane determines the type of
$\genh$. \label{enhancef}}
\end{table}

\section{Finite Enhancement: $f(\mz ) < 1$}

Let us explore how the value of $f(\mz)$ determines the enhanced
algebra.  First, consider the values $f(\mz) < -1$; it is easy to see
that equation (\ref{ll}) cannot be satisfied simultaneously with
$\mJ^2 \geq -2$.  Thus no new roots will appear, and we merely find
$\genh = \gg \oplus u(1)$.

At the saturating value $f(\mz) = -1$,
\begin{eqnarray}
\ll = -\mJ^2 - 2 n^2 \,,
\end{eqnarray}
which only has solutions for $\mJ^2 = -2$, with $\ll = 0$ and $n = \pm
1$.  The new simple root is $\malpha_0 = (n_0, \lambda) = (1, 0)$, and
satisfies $(\malpha_0, \malpha_i) = 0$ for all $\malpha_i$ in $\gg$.
Thus the enhancement is $\genh = \gg \oplus A_1$, independent of $\gg$.

Now consider the range $-1 < f(\mz) <1$, or equivalently, where the
coefficient $(f(\mz)-1) $ of $n^2$ in (\ref{ll}) is negative.
It follows that this equation only has solutions for $\mJ^2 = -2$, all
of which are roots of $\genh$ with $\LL = 2$.  Moreover the number of
new roots is finite, for (\ref{ll}) will have no solution for
sufficiently large $|n|$. Therefore $\genh$ is a finite Lie algebra.  We

now give several examples of this case.

Consider the ${\bf H_{N}}$ series, representing the algebra $A_{N}$ on
branes $\mA^{N+1} \mC$.  The $\mC$-brane enables the configuration of
otherwise mutually local branes to have junctions with asymptotic
\mbox{$q$-charge} , thus making it possible for new roots to stretch to
an enhancing $[p,q]$ brane when it is added.  The charge quadratic
form is (see table 1)
\begin{eqnarray}
f(p,q) = -\,\frac{1}{N+1}\; (p^2 - (N+3)pq + (2N+3)q^2) \,,
\end{eqnarray}
and will determine what kind of enhanced algebra appears.

The simplest enhancement occurs when the new brane is another $[1,0]$,
\mbox{$f(1,0) = -1/(N\!+\!1)$}.  The only junctions satisfying
\myref{ll} besides the roots of $A_{N}$ have $|n|=1$ and
$\ll=N/(N\!+1\!)$, requiring either $\lambda \in {\bf N+1}$
(fundamental) or $\lambda \in {\bf \overline{N+1}}$ (antifundamental).
We can choose $\malpha_0 = (1, \lambda)$ with $\lambda = - \omega^1$
or $\lambda = - \omega^{N}$. In either case we get a total of $2(N+1)$
new roots from $n= \pm 1$, the number of roots needed to enhance from
$A_{N}$ to $A_{N+1}$.  Indeed, $(\malpha_0, \malpha_i) = -( \momega^1,
\malpha_i) = \delta^1_i$ or $(\malpha_0, \malpha_i) = - (\momega^{N},
\malpha_i) = \delta^{N}_i$, and either choice reproduces the Cartan
matrix and Dynkin diagram of $A_{N+1}$.  As expected, we enhance to
${\bf H_{N+1}}$.

\begin{table}[t]
\begin{center}
\begin{tabular}{|c|c|c|c|c|c|} \hline
$\rule{0mm}{5mm}\gg$ & Config &Branes & $f(\mz)$ & $\genh$ &
Enh. branes \\
\hline \hline
 &&  & $-1/N$ & $A_{N+1}$ & $\mA{\bf H_N} $  \\
$A_N$ & ${\bf H_N}$ & $\mA^{N+1} \mC$ & $(N\hskip-2pt -\hskip-2pt
3)/(N\hskip-2pt
+\hskip-2pt 1)$ &
$D_{N+1}$ &
${\bf H_N}\mX_{\bf [3,1]}$ \\
 & &  & $(2N\hskip-2pt -\hskip-2pt 7)/(N\hskip-2pt +\hskip-2pt 1)$
 & $E_{N+1}$ & ${\bf H_N} \mX_{\bf [4,1]}$
\\
\hline
$D_N$ & ${\bf D_N}$& $\mA^N \mB \mC$ & $0$ & $D_{N+1}$ & $\mA{\bf D_N}$
\\
  &&   &  $(N-4)/4$ & $E_{N+1}$ & ${\bf D_N} \mC$ \\ \hline
$A_1$ &${\bf E_1}$& $\mB\mC\mC$ & $a^2/2 - 1$ & $\tilde{A_2}^{(a)}$ &
${\bf E_1}\mX_{\bf [2a-1,1]}$ \\ \hline
$\rule{0mm}{5mm}D_{4}$ & ${\bf D_{4}}$ & ${\bf A}^{4}{\bf BC}$ & $0$ &
$D_{5}$ & ${\bf D_4} \mX_{\bf [p,q]}$ \\ \hline
 & & & $0$ & $D_6$ & $\mA{\bf D_5}$\\
$D_5$ &${\bf D_5}$ & $\mA^5\mB\mC$ & $1/4$ & $E_6$& ${\bf D_5}\mC$  \\
 & $\cong {\bf E_5}$&& $1$ & $\widehat{D}_5$ & ${\bf D_5} \mX_{\bf
[1,2]}
$\\
&& & $9/4$ & $ E_4^H$ & ${\bf D_5} \mX_{\bf [1,3]}$ \\\hline
  &   && $1/3$ & $E_7$& $\mA{\bf E_6}$\\
$E_6$ &${\bf E_6}$& $\mA^5\mB\mC\mC$    & $1$ & $\widehat{E}_6$ & ${\bf
E_6}\mX_{\bf
[3,1]}$\\
     &   && $7/3$ & $ E_5^H$& ${\bf E_6} \mX_{\bf [4,1]}$\\ \hline
     &   && $1/2$ & $E_8$& $\mA{\bf E_7}$\\
$E_7$ &${\bf E_7}$  & $\mA^6\mB\mC\mC$ & $1$ & $\widehat{E}_7$&
${\bf E_7}\mX_{\bf [3,1]}$\\
   &     && $5/2$ & $ E_6^H$ & ${\bf E_7} \mX_{\bf [4,1]}$ \\ \hline
$E_8$ &${\bf E_8}$   & $\mA^7\mB\mC\mC$     & $1$ & $\widehat{E}_8 =
E_9$
& $\mA{\bf E_8}$\\
  & && $3$ & $E_7^H$& ${\bf E_8} \mX_{\bf [4,1]}$\\ \hline
\end{tabular}
\end{center}
\caption{Finite algebras $\gg$ and enhanced
algebras $\genh$ obtained by adding a
single $\mz = [p,q]$ brane.  The Dynkin diagram of $\tilde{A}_2^{(a)}$
has two nodes  with $a$ lines joining them.
\label{enhancements}}
\end{table}

Another possibility that proceeds identically for any algebra in the
series is to add another $[1,1]$ brane.  We would expect that the pair
of $\mC$-branes will now form an additional \mbox{$A_{1}$ algebra},
giving $A_{N} \oplus A_{1}$.  Indeed, as $f(1,1) = -1$ for all $N$,
and, as mentioned above, this simply adds an additional $A_{1}$
algebra.

Other enhancements are possible.  For example, adding a $[3,1]$-brane
gives $f(3,1)\!\!=\!\!(N\!-\!3)\, /\, (N\!+\!1)$, so at $n^2=1$, $\ll
= 2(N-1)/(N+1)$.  This is the Weyl orbit with dominant weight
$\omega^2$ (or conjugate) and thus the new simple root is $\malpha_0 =
(1,-\omega^2)$.  The inner products of this with the other simple
roots show the enhancement $A_{N} \ra D_{N+1}$.  We can pass the
$[3,1]$-brane through the $[1,1]$ branch cut to turn it into a
$[1,-1]$-brane, thus recovering the canonical form of the ${\bf D_N}$
series. Analogously, adding a $[4,1]$-brane enhances to the
\mbox{$E_N$ series}.  In general a $[p,1]$-brane adds a node
connecting to the $(p-1)^{th}$ node of the $A_{N}$ Dynkin diagram.

Naturally we do not need to begin with the ${\bf H_N}$ series.  One can
consider enhancing the ${\bf D_N}$ or ${\bf E_N}$ series as well.  A
particularly interesting example is $D_4$, on $\mA^4 \mB \mC$.  For this

case $f(p,q)$ vanishes identically, and thus the enhanced algebra
$\genh$ is
independent of the charges $\mz = [p,q]$ of the enhancing brane $\mZ$.
This is easy to understand: the monodromy of $D_4$ is minus the
identity matrix and is invariant under the $\sl2z$ transformation
relating any two choices for new branes.  We have $\malpha_0 = (1,
\lambda)$ with $\ll = 1$, and therefore $\malpha_0 = (1, -\omega^i)$
where $\omega^i = 1,3,4$ is the highest weight of the ${\bf 8_v}$,
${\bf 8_s}$ or ${\bf 8_c}$ representation.  For any of these three
choices, the enhanced algebra is $\genh = D_{5}$, a manifestation of
triality.

The various $D_N$ algebras can enhance to $E_{N+1}$ algebras.  For
$D_N$, $f(p,q) = (N-4) \, q^2/4$, and the value of $p$ does not affect
the resulting enhancement.  As an example, $q = 1$ gives $f(\mz) =
(N-4)/4$, which means the new roots $(1, \lambda)$ have $\ll = n/4$.
Such weights belong to the spinor representations and enhancement
proceeds by attaching a new node to a node associated to a spinor
representation, the result being $E_{N+1}$. If $p=1$, $E_{N+1}$ is
obtained in the canonical form; otherwise a global $\sl2z$
transformation is necessary to recover the usual monodromy.

If we restrict ourselves to $f(\mz) < 1$ it is simple to show that
$E_6$ can only enhance to $E_7$, that $E_7$ can only enhance to $E_8$
and that $E_8$ cannot enhance to any finite algebra. This last fact
follows simply because for $E_8$ one has $f(\mz) \geq 1$ for any
choice of $\mz \not= 0$.

As $N$ increases in the ${\bf H_N}$, ${\bf D_N}$ and ${\bf E_N}$
series, the values of $f(\mz)$ tend to grow larger.  When an enhancing
brane has charges giving $f(\mz) \geq 1$, the algebras that result
will not be finite-dimensional, but infinite-dimensional.  We shall
study these cases in \cite{infinite}.


\section{$\sl2z$ Conjugacy Classes and Classification}
\label{classify}

Having explored the appearance of various algebras on 7-branes in the
previous sections, we now proceed to find their place in the
classification scheme.  As mentioned in the introduction, the complete
classification involves a discussion of infinite dimensional Lie
algebras. Therefore, a complete analysis will be postponed for the
sequel paper \cite{infinite}.  Here we will present the complete table
of results, including information to be obtained in the sequel, but
only parts of this table will be explained.

The monodromy $K$ of a brane configuration is the primary factor
determining the associated algebra.  Global $\sl2z$ transformations
organize monodromies into equivalence classes which are physically
distinct.  Thus it is the conjugacy classes of elements of the modular
group $\sl2z$, each of which corresponds to an equivalence class, that
we must study.

Given that the trace is a conjugation invariant, conjugacy classes of
$\sl2z$ are conveniently organized according to the value of the
trace.  An element $K\in \sl2z$ is called {\it elliptic} if $|{\Tr}
K|<2$, {\it parabolic} if $|{\Tr} K|=2$ and {\it hyperbolic} if
$|{\Tr} K|>2$. Under the action of $\sl2z$, elliptic elements have one
fixed point in the upper half plane, whereas the fixed points of the
parabolic and hyperbolic elements are real rational and irrational
numbers respectively.  We are curious how many conjugacy classes exist
at a given value of $t \equiv {\Tr} K$, each of which corresponds to
an inequivalent 7-brane configuration.  For the case of elliptic and
parabolic monodromies the number of conjugacy classes $H(t)$ can
determined by elementary methods.  Let us discuss these cases.

{\bf Elliptic conjugacy classes:} Consider an element $K\in \sl2z$
with ${\Tr} K = 0$.  The characteristic equation of $K$ implies that
$K^{2}=-\unit$.  The fixed point of $K$ is thus left invariant by a
cyclic
group of order two. Its image on the fundamental domain ${\cal F}$ of
the modular group must be $z= i$, for this is the unique point in
${\cal F}$ left invariant by a group of order two, the group generated
by $S$.  Since $S$ and $-S (= S^{-1})$ act in the same way in the
upper half plane and can readily be shown not to be equivalent
matrices, we can only have that $gKg^{-1}\in\{S,S^{-1}\}$. Thus
trace zero elements fall into two conjugacy classes.

If ${\Tr} K =1$ then the characteristic equation requires
$K^{3}=-\unit$, and the fixed point of $K$ is left invariant by a
group of order three. Its image in ${\cal F}$ must be either $\exp(\pi
i/3)$ or $\exp (2\pi i/3)$, both of which have isotropy groups of
order three.  There groups are generated by $TS$ and $ST$
respectively.  It is readily verified that $ST$ is the relevant
generator and we get two inequivalent classes, that is, $gKg^{-1}
\in\{ST,(ST)^{-1}\}$. When ${\Tr} K = -1$ a completely analogous
argument gives the classes $\{ -ST, -(ST)^{-1}\}$.

{\bf Parabolic conjugacy classes:} If ${\Tr} K= \pm 2$, then $K$ is of
infinite order and has a real rational fixed point. This point can be
mapped to infinity by an $\sl2z$ transformation $g$. Then infinity is
a fixed point of $gKg^{-1}$. The only elements of $\sl2z$ that have
infinity as a fixed point are of the type
$\pm\!\left({1\;N\atop0\;1}\right)$ A simple computation shows that
none of these matrices are conjugate in $\sl2z$. Thus $gKg^{-1}\in \{
\pm\!\left({1 \; N \atop 0 \; 1}\right) | N \in \bbbz \}$, and
elements of trace plus or minus two have infinitely many conjugacy
classes.

\medskip
The hyperbolic conjugacy classes have fixed points which are
irrational real numbers, which cannot be mapped to ${\cal F}$.  As a
result, enumerating these classes is a more difficult problem, and
requires other methods.  In fact, $H(t)$ can be determined using the
isomorphism between the $\sl2z$ matrices of trace $t$ and binary
quadratic forms of discriminant $t^{2}-4$, discussed in section
\ref{binary}. It is clear from the isomorphism that the number of
conjugacy classes for trace $t$ and trace $-t$ are equal.  Values of
$H(t)$ for $|t| \leq 7$ are listed in Table~\ref{maintable}.  We
explain how to calculate $H(t)$ for generic $t$ in an appendix of
\cite{infinite}.

We now wish to organize the brane configurations we have discussed
throughout the paper into the appropriate conjugacy classes.  We must
note first that there are certain brane configurations which have $K =
\unit$, and so are invisible to the total monodromy.  If a set of
branes with $K = \unit$ is added to some configuration $\mG$, the
resulting configuration $\mG'$ will have $K(\mG') = K(\mG)$.  Were
these configurations very common, our classification would be
hopeless.  It can be proven, however, that the number of branes in a
configuration with unit monodromy must be a multiple of 12. Thus any
monodromy realized on $n$ branes will also have realizations on $n +
12k$ branes, for any positive integer $k$.  Specifying the number of
branes as well as the monodromy fixes this ambiguity. As will be
discussed in \cite{infinite}, a configuration with twelve branes and
unit monodromy realizes the infinite dimensional loop algebra
$\widehat{E_{9}}$ algebra.  Below we shall be assuming each
configuration has fewer than twelve 7-branes.

In \secref{monodet}, we proved that if
$\gg$ is an algebra on a brane configuration $\mG$ with non-degenerate
Cartan matrix $A(\gg)$, then
\begin{equation}
\mQ^2\;\mbox{det}A(\gg) = 2-t.
\label{maineq}
\end{equation}

This equation will be a useful tool in classifying the various $\mG$.
Let us start by analyzing configuration of 7-branes with monodromy of
trace zero.  It follows from (\ref{maineq}) that det$A(\gg) = 2$,
implying that the possible finite simple algebras are $\gg=\{A_1,
E_7\}$.  The possible conjugacy classes at this trace are represented
by $\{S, -S\}$.  Comparing with the configurations we are familiar
with, we see that $K({\bf E_7})$ is conjugate to $S$, and that the
${\bf H_1}$ configuration, realizing the algebra $A_1$, has monodromy
conjugate to $(-S)$, exhausting these two conjugacy classes.

Now consider the case when the collection of 7-branes has ${\Tr}K =
-1$.  There are this time the classes $\{ -ST, -(ST)^{-1}\}$.  We see
from (\ref{maineq}) that det $A(\gg)=3$, and thus the candidate
algebras are $\gg=\{ A_{2},E_{6}\}$.  In fact the monodromies show
that $(-ST)$ is associated to ${\bf H_2}$, which realizes $A_2$, and
$(-(ST)^{-1})$ is associated to the ${\bf E_6}$ configuration.

The case where ${\Tr} K = 1$ requires det$(A(\gg)) = 1$; the only such
$ADE$ algebra is $E_8$, which is realized by the conjugacy class
$ST$.  The class $(ST)^{-1}$, on the other hand, corresponds to the
${\bf H_0}$ configuration, which does not support an algebra.

The information we have just derived for the elliptic conjugacy
classes is summarized in Table \ref{maintable} which can be viewed as
the extension of Table I of \cite{Kodaira}.  Notice that
(\ref{maineq}) required $\ell =1$ for all these cases.

Both the ${\bf E_N}$ and ${\bf H_N}$ configurations extend naturally
to all $N\geq 0$.  The ${\bf E_N}$ configurations all satisfy the
relation
\begin{eqnarray}
\label{etrace}
{\Tr} K({\bf E_N}) = N - 7 \,.
\end{eqnarray}
At ${\Tr} K = -6$ both inequivalent realizations ${\bf E_1}$ and ${\bf
\widetilde E_1}$ discussed in section 2 appear, and at ${\Tr} K = -7$
we have ${\bf \tilde{E}_0}$. For ${\bf E_{N}}, N\ge 2$, we have det
$A({E_N}) = 9 - N$ and $\mQ({\bf E_{N\ge 2}})=1$, while for ${\bf
E_{1}}$, det $A({\bf E_1}) = 2$ and $\mQ({\bf E_1})=2$, and for ${\bf
\tilde{E}_1}$, det $A({\bf \tilde{E}_1}) = 8$ and $\ell({\bf
\tilde{E}_1}) = 1$; (\ref{etrace}) then follows from (\ref{maineq}).
The configurations ${\bf E_2}$ and ${\bf \tilde{E}_1}$ realize the
non-semisimple algebras $A_1 \oplus u(1)$ and $u(1)$, so the
intersection matrix $A$ is not a Cartan matrix.

The junction lattices of configurations of two mutually nonlocal
branes such as ${\bf \tilde{E}_0}$ consist solely of the asymptotic
charge parts.  Thus there is no algebra, and (\ref{maineq}) is
modified to just $\ell^2 = 2 - t $, as derived directly in 
\myref{2trace}.
For ${\bf \tilde{E}_0}$ we find $\ell = 3$, as required.

For $N\geq 9$ the ${\bf E_N}$ series gives
infinite-dimensional algebras.  The ${\bf E_N}$ configurations that
correspond to elliptic conjugacy classes correspond to Kodaira
singularities and are collapsible, as is the parabolic ${\bf E_5}
\cong {\bf D_5}$.

For all positive values of $N$, the ${\bf H_N}$ configurations realize
an $A_N$ algebra.  The traces satisfy
\begin{eqnarray}
\label{htrace}
{\Tr} K({\bf H_N}) =  1-N \,,
\end{eqnarray}
in all cases consistent with (\ref{maineq}), making use of $\ell({\bf
H_N}) = 1$.  We notice that only ${\bf H_{N}}$ brane configurations
associated to elliptic conjugacy classes are collapsible,
indeed ${\bf H_3} \cong {\bf D_3}$, is not a Kodaira singularity.
${\bf H_0}$ has $\ell = 1$, satisfying (\ref{maineq2}).

More exotic brane realizations of $A_N$ exist, the ${\bf A_N^{{\bf
i}\,\ell}}$
series, characterized by the values of $i$ and $\ell$, as discussed in
section 
\ref{kodaira}.  The value $\ell = 1$ gives ${\bf H_N}$.  These have
${\Tr} K = 2 - \ell^2 (N+1)$, which satisfies (\ref{maineq}).  They
are beyond the range of Table \ref{maintable}, with the exception of
$N=1$, $\ell = 2$, which is just equivalent to ${\bf E_1}$.  It
is interesting that ${\bf E_1}$, which is the only member of the
${\bf E_N}$ series not equivalent to a member of ${\bf S_{p=2,N}}$,
turns up as equivalent to ${\bf A_1^{{\bf i}=1,\ell = 2}}$.

Let us now consider the 7-brane configurations with ${\Tr}K = -2$. It
follows from (\ref{maineq}) that det $A(\gg)= 4$ and $\ell = 1$, or
det $A(\gg) = 1$ and $\ell = 2$.  Therefore the possible finite
algebras are $\gg=\{D_{N}, E_8\}$.  All $D_N$ are possible since det
$A({D_N}) = 4$ holds for each one.  This conveniently coincides with
the parabolic conjugacy classes, which we know are infinite in number.
We do not observe an $E_8$ configuration with $\ell =2$.  There is a
unique configuration of two seven branes with trace minus two, which
has no algebra, the $\mB\mC$ configuration recognized as ${\bf D_0}$;
it has $\ell = 2$ and so satisfies (\ref{maineq2}).  Each member of
the series of ${\bf D_N}$ algebras is obtained from ${\bf D_0}$ by
adding $N$ $\mA$-branes, and they are all characterized by
\begin{eqnarray}
{\Tr} K({\bf D_N}) = -2 \,.
\end{eqnarray}
Of the
infinitely many conjugacy classes of trace minus two, only
those with representatives of the type
\begin{eqnarray}
\pmatrix{-1&N-4\cr 0&-1}, \quad N\geq 0 \,,
\end{eqnarray}
are realized by the $D_N$ algebras. Other conjugacy classes are
occupied by infinite dimensional algebras \cite{infinite}.
The ${\bf D_N}$ configurations with $N \geq 4$ are collapsible.

Finally, configurations of branes with ${\Tr}K = 2$ include the
straightforward series of mutually local D7-branes ${\bf A_N} =
\mA^{N+1}$, realizing the algebra $A_N$:
\begin{eqnarray}
{\Tr} K({\bf A_N}) = 2 \,.
\end{eqnarray}
As mentioned in section 3, $\ell=0$ for these
configurations and their Cartan matrices
must satisfy (\ref{maineqA}), which they do.  They are all
collapsible.  We noticed above that ${\bf E_9}$ also occurs at ${\Tr}
K = 2$, and in fact the entire affine exceptional series is present at
this trace as will be explored in \cite{infinite}.

The data we have assembled in the table provides evidence that the
monodromy of configurations of 7-branes, together with the number of
branes and the asymptotic charge constraint $\ell$, determines the
algebra realized on the configuration.  With one exception, we find
that for each conjugacy class of $\sl2z$ a single non-trivial algebra
is realized on the configuration with the minimum number of branes.
Configurations corresponding to elliptic conjugacy classes
with minimal numbers of branes  
are collapsible, and correspond to Kodaira singularities.  This is also
the case for some of the parabolic conjugacy classes, of which there
are an infinite number, but not for others.  None of the hyperbolic
conjugacy classes correspond to collapsible configurations.

\begin{table}
\begin{center}
\begin{tabular}{|c|c|c|c|l|c|}
\hline
$\rule{0mm}{5mm}$ $K$-type & ${\Tr}K$ & det $A(\gg)$ &
   Brane configuration $\mG$ & $\hspace{.35in}K$ & $H(t)$ \\ \hline
\hline
&$\rule{0mm}{7mm}$$-7$ &9& $ {\bf \tilde{E}_0}\,,{\bf H_8} $
&$\pmatrix{-7 & 1\cr -1& 0}^{\pm1}   $  & 2
\\ \cline{2-6}
&$\rule{0mm}{7mm}$$-6$ &(2,8),8& $ ({\bf E_1}, {\bf \tilde{E}_1})
                                        \,, {\bf H_7}  $
&$\pmatrix{-6 & 1\cr -1& 0}^{\pm1}  $  & 2
\\ \cline{2-6}
$hyp.$
&$\rule{0mm}{7mm}$$-5$ & 7 & $ {\bf E_2}\,, {\bf H_6}  $
&$\pmatrix{-5 & 1\cr -1& 0}^{\pm1}  $  & 2
\\ \cline{2-6}
&$\rule{0mm}{7mm}$$-4$ & 6 & $ {\bf E_3}\,, {\bf H_5}  $
&$\pmatrix{-4 & 1\cr -1& 0}^{\pm1}  $  & 2
\\ \cline{2-6}
&$\rule{0mm}{7mm}$$-3$ & 5 & $ {\bf E_4}  = {\bf H_4}  $
&$\pmatrix{-3 & 1\cr -1& 0}  $  & 1
\\ \hline
$par.$
&$\rule{0mm}{8mm}$$-2$ &4  &
$\begin{array}{c} {\bf D_{N+4 \,\geq 0}} \\
 ({\bf E_5}\!=\!{\bf D_5}\,,{\bf H_3}\!=\!{\bf D_3})\end{array}$
&$ \pmatrix{-1 & N\cr  0&-1} $  & $\infty$
\\ \hline
&$\rule{0mm}{5mm}$$-1$& 3&${\bf E_6}\,,{\bf H_2}$
&$  ~-\!(ST)^{\mp1}$&2
\\ \cline{2-6}
$ell.$
&$\rule{0mm}{5mm}$ 0& 2&${\bf E_7}\,,{\bf H_1}$
&$  ~~~~~~S^{\pm1}    $&2
\\ \cline{2-6}
&$\rule{0mm}{5mm}$ 1& 1&${\bf E_8}\,,{\bf H_0}$
&$ ~~~~(ST)^{\pm1}   $&2
\\ \hline
$par.$
&$\rule{0mm}{8mm}$ 2 & $\begin{array}{c} N\\ 0\end{array}$
&$\begin{array}{c} {\bf A_{N-1\,\geq 0}}\\
           {\bf \widehat{E}_{N+9\geq 0}},{\bf\widehat{\tilde{E}}_{1}},
           ({\bf E_9}={\bf \widehat{E}_8}) \end{array}$
& $\pmatrix{ 1 &-N\cr  0& 1}$ & $\infty$
\\ \hline
&$\rule{0mm}{7mm}$ 3 &  $-1 $& $ {\bf E_{10}}  = {\bf E_8^H}$
& $\pmatrix{ 0 & 1\cr -1&3 }$ & 1
\\ \cline{2-6}
&$\rule{0mm}{7mm}$ 4 & $ -2 $& $ {\bf E_{11}}\,, {\bf E_7^H}$
& $\pmatrix{ 0 & 1\cr -1&4 }^{\pm1}$ & 2
\\ \cline{2-6}
$hyp.$
&$\rule{0mm}{7mm}$ 5 & $ -3 $& $ {\bf E_{12}}\,, {\bf E_6^H}$
& $\pmatrix{ 0 & 1\cr -1&5 }^{\pm1}$ & 2
\\ \cline{2-6}
&$\rule{0mm}{7mm}$ 6 & $ -4 $& $ {\bf E_{13}}\,, {\bf E_5^H}$
& $\pmatrix{ 0 & 1\cr -1&6 }^{\pm1}$ & 2
\\ \cline{2-6}
&$\rule{0mm}{7mm}$ 7 & $ -5 $& $ {\bf E_{14}}\,, {\bf E_4^H}$
& $\pmatrix{ 0 & 1\cr -1&7 }^{\pm1}$ & 2 \\ \hline
\end{tabular}
\end{center}
\caption{$\sl2z$ conjugacy classes and algebras realized on 7-branes
with overall monodromy $K$ (up to conjugation). The upper and lower
exponents of the matrices correspond to the first and the second brane
configuration, respectively. The brane configurations $\mG$, whose
notations suggest the algebra $\gg$, are defined in the text. The
determinant of $A(\gg)$ is given, except when no algebra is realized,
as with ${\bf H_0,\tilde{E}_0}$ and ${\bf D_0}$.  \label{maintable}}
\end{table}

This paper has restricted its attention mostly to finite algebras on
7-branes. To fully understand the possibilities one should also
consider the infinite-dimensional algebras which appear on
non-collapsible configurations of 7-branes, and are associated to
certain parabolic and hyperbolic conjugacy classes. These will be
explored in \cite{infinite}, where we consider the affine and
hyperbolic extensions of the exceptional series, the
infinite-dimensional $E_N$ algebras with $N \geq 10$, and others.
Some of the relevant algebras are not Kac-Moody and do not possess a
Cartan matrix, but instead are understood as loop algebras of other
infinite-dimensional algebras.  A particularly important example is
$\widehat{E}_9$, the algebra associated to the simplest configuration
of unit monodromy.  Taken together, a fascinating pattern of algebraic
enhancements on 7-brane configurations emerges.

\subsection*{Acknowledgments}

Thanks are due to {\bf A}.~Hanany, {\bf
D}.~Zagier and {\bf E}.~Witten for their useful comments and
questions.  We are grateful to C.~Vafa for instructive discussions,
and to V.~Kac and R.~Borcherds for very helpful correspondence.

This work was supported by the U.S.\ Department of Energy under
contract \#DE-FC02-94ER40818.

\newpage
\section*{Appendix}
In this appendix we prove two equations relating the determinant of
the metric on the lattice of junctions and that on the lattice of
junctions with zero asymptotic charges to the trace of the overall
monodromy.

\subsection*{Trace-Determinant relation \#1}

We first compute the determinant of the total intersection matrix
(defined in \cite{dewolfezwiebach}), which is the metric on the space of

junctions. Thus we are interested in:
\begin{equation}
\mbox{det} \A \equiv
\left|\begin{array}{cccc}
1 & a_{12} & \ldots &a_{1n} \\
a_{12} & 1 & \ldots &a_{2n} \\
\vdots & \vdots & \ddots & \vdots \\
a_{1n} & a_{2n}& \ldots &1
\end{array}\right| \hspace{.5in}
a_{ij} = -\frac{1}{2}(\mathbold{z}_i\times \mathbold{z}_j)=
-\half(p_iq_j-q_ip_j).
\label{basisdef}
\end{equation}
Consider the 7-brane configuration with charges
$\{\mz_1\!=({p_1\atop q_1}),\ldots, \mz_n\!=({p_n\atop q_n})\}$. We
claim that the following relation holds between $\mbox{det} \A$ and
the trace of the overall monodromy:
\begin{eqnarray}
\begin{array}{|c|} \hline \\
\mbox{det}\A = \frac{1}{4} {\Tr} K+\frac{1}{2} \\
\\ \hline\end{array}
\label{mystery}
\end{eqnarray}
Both sides of \myref{mystery} are functions of the $(p,q)$-charges and
are explicitly given, thus we need to prove an algebraic identity.
In particular ${\Tr} K$ can be expressed in terms of $p_i$ and $q_i$
using \myref{threemon}:
\begin{eqnarray}
{\Tr} K= 2 + \sum_{i_1<i_2}
(\mz_{i_1}\!\!\times\!\mz_{i_2})
(\mz_{i_2}\!\!\times\!\mz_{i_1}) + \ldots +
\sum_{i_1<\ldots<i_n}
(\mz_{i_1}\!\!\times\!\mz_{i_2})\ldots
(\mz_{i_n}\!\!\times\!\mz_{i_1}).
\label{barton}
\end{eqnarray}
The proof of \myref{mystery} goes as follows. Notice that both the lhs
and the rhs is a quadratic polynomial in each $q_i$ for any fixed
values of the $p_i$'s. As a consequence it is sufficient to verify
\myref{mystery} at three distinct values of $q_i$ for each $i$ (i.e.
at $3^n$ points) while keeping the $p_i$'s arbitrary. The natural
choice for these points is $-1, 0$ and 1; in other words we should
verify \myref{mystery} for brane configurations $\{\mz_1\!=({p_1\atop
q_1}),\ldots, \mz_n\!=({p_n\atop q_n})\}$ where the $q$-charge of each
brane is $-1, 0$ or 1 (but one has to include $({p\atop 0})$ branes for
$|p|>1$ as well). The $({p\atop -1})$ branes can be turned into
$({-p\atop 1})$-branes and the $({p\atop 0})$-branes can be moved to the

left of the configuration by pulling all the other branes through
their branch cut; note that this transformation does not change their
$q$-charge which remains $1$. Therefore to show that \myref{mystery} is
true in general, {\em it is sufficient to prove its validity for brane
configurations of the form}
\begin{equation}
\{\mz_1\!=\left({p_1\atop 0}\right),\ldots,
\mz_k\!=\left({p_k\atop 0}\right),
\mz_{k+1}\!=\left({p_{k\!+\!1}\atop 1}\right),\ldots,
\mz_n\!=\left({p_n\atop 1}\right)\}.
\label{simpleconf}
\end{equation}

As the next step, we calculate $\mbox{det}\A$ for the above
configuration. Let us expand the determinant as follows:
\bea
\label{bigdet}
\mbox{det}\A &=& 1 +
\sum_{i_1<i_2}\mbox{det}_2(\mz_{i_1}\mz_{i_2})+\ldots+
\sum_{i_1<\ldots<i_m}\mbox{det}_m(\mz_{i_1}\ldots \mz_{i_m}) \\
\mbox{det}_m(\mz_1\ldots \mz_m) &\equiv&
\left|\begin{array}{ccccc}
0 & a_{12} & a_{13}& \ldots &a_{1m} \\
a_{12} & 0 & a_{23}& \ldots &a_{2m} \\
a_{13} & a_{23} & 0& \ldots &a_{3m} \\
\vdots & \vdots & \vdots & \ddots & \vdots \\
a_{1m} & a_{2m}& a_{3m}& \ldots &0
\end{array}\right|,
\end{eqnarray}
where we collected the terms containing $m$ factors from the
diagonal. We want to compute $\mbox{det}_m$ for the configuration
\myref{simpleconf}, in which case
\begin{equation}
-2a_{ij} = \left\{\begin{array}{lll}
0       & \mbox{if} &i,j\le k \\
p_i     & \mbox{if} & i\le k;j>k \\
p_i-p_j & \mbox{if} & i,j>k.
\end{array}\right.
\end{equation}
When $k\ge2$, the first two lines of $\mbox{det}_m$ are proportional
and the determinant is zero. Let us consider the case $k=0$,
first. Elementary operations on the determinant give:
\bea
\mbox{det}_m =
\left|\begin{array}{cccccc}
0 & a_{12} & a_{13}& \ldots &a_{1m} \\
a_{12} & a_{21} & a_{21}& \ldots &a_{21} \\
a_{23} & a_{23} & a_{32}& \ldots &a_{32} \\
\vdots & \vdots & \vdots & \ddots & \vdots \\
a_{m-1,m} & a_{m-1,m}& a_{m-1,m}& \ldots &a_{m,m-1}
\end{array}\right|
=
\left|\begin{array}{cccccc}
0 & a_{12} &a_{13}& \ldots &a_{1n} \\
a_{12} & 0 & 0&  \ldots &  0 \\
a_{23} & 2a_{23} & 0& \ldots &0 \\
\vdots & \vdots & \ddots  &  &\vdots \\
a_{n-1,n} & 2a_{n-1,n}& \ldots &2a_{n-1,n}& 0
\end{array}\right|. \nn
\end{eqnarray}
To obtain the first form we subtracted the $m-1$th row from the $m$th,
then the $m-2$th from the $m-1$th \ldots and finally the 1st from the
2nd. Then we obtained the rhs by adding the 1st column to all the
others. The final form yields:
\begin{eqnarray}
\mbox{det}_m(\mz_1\ldots \mz_m)=
(-)^m2^{m-2}(a_{12}a_{23}\ldots a_{m1})=
\frac{1}{4}
(\mz_1\!\!\times\!\mz_2)\ldots
(\mz_m\!\!\times\!\mz_1).
\label{subdet}
\end{eqnarray}
When $k>1$, \myref{subdet} is trivially satisfied. Now consider $k=1$,
i.e. when the leftmost brane has vanishing $q$-charge, the others have
$q_i=1$. Manipulations similar to the $k=0$ case yield:
\begin{eqnarray}
\left|\begin{array}{cccccc}
0 & p_1 & p_1& p_1& \ldots & p_1 \\
p_1 & 0 & a_{23}& a_{24}& \ldots &a_{2m} \\
p_1 & a_{23} &  0& a_{34} & \ldots &a_{3m} \\
p_1 & a_{24} &a_{34} &  0  & \ldots &a_{3m} \\
\vdots & \vdots & \vdots & \vdots & \ddots & \vdots \\
p_1 & a_{2m}& a_{3m}& a_{4m}&  \ldots &0
\end{array}\right|
=
\left|\begin{array}{cccccc}
0 & p_1 & p_1 & p_1 & \ldots & p_1 \\
p_1 & 0 & a_{23}& a_{24}&\ldots & a_{2n} \\
0 & a_{23} & 0 & 0 & \ldots &0 \\
0 & a_{34} & 2a_{34} & 0&\ldots &0\\
\vdots & \vdots & \vdots&\vdots & \ddots & \vdots \\
0 & a_{n-1,n}& 2a_{n-1,n}& 2a_{n-1,n}& \ldots &0
\end{array}\right|,
\end{eqnarray}
which after expanding the determinant with respect to the first column
and then the last column yields again \myref{subdet}.
Comparing \myref{subdet}, \myref{bigdet} and \myref{barton}
proves the claimed equality \myref{mystery}.

\subsection*{Trace-Determinant relation \#2}

Consider the lattice of junctions $\Lambda$ supported by a 7-brane
configuration consisting of $n$ branes. Let us set one of the branes
to be a $[1,0]$-brane by an $\sl2z$-transformation, then the smallest
values of asymptotic charges, $|p|$ and $|q|$ are $1$ and $\mQ$,
respectively. The junction lattice is $n$-dimensional, the metric in
the particular basis furnished by open strings supported on a single
7-brane is given by \myref{basisdef} and the volume of the unit cell
of $\Lambda$ (which is of course basis-independent) is
$\sqrt{\mbox{det}\A}$. We are interested in the sub-lattice
$\Lambda_0\subset\Lambda$ which is generated by junctions of zero
asymptotic charges and want to compute the volume of its unit cell. We
have to distinguish two cases: when the 7-branes are all mutually
local $\Lambda_0$ is $r=n-1$ dimensional while for mutually nonlocal
branes it is $r=n-2$ dimensional. Moreover, as we will see,
the case of $\mbox{Tr}K=2$ has to be treated separately for mutually
nonlocal branes. 

{\bf Mutually local branes.} In this case $K=\mbox{$({1 \; -n \atop 0
\;\;\; 1 } )$},\;\mbox{Tr}K=2$ and it is straightforward to find an
explicit basis on $\Lambda_0$:
\begin{eqnarray}
\Lambda_0 = \{\alpha_i\}_{i=1}^r,\;\;\;\;\;\;
\alpha_i = {\bf x}_i-{\bf x}_{i+1},
\end{eqnarray}
where ${\bf x}_i$ are the basis strings having unit support on the $i$th
7-brane only. The determinant is readily computed:
\begin{eqnarray}
\left|
\begin{array}{ccccc}
-(\alpha_1,\alpha_1)&\ldots&-(\alpha_1,\alpha_r) \\
\vdots&\ddots&\vdots   \\
-(\alpha_r,\alpha_1)&\ldots&-(\alpha_r,\alpha_r) 
\end{array}
\right| &=& 
\left|
\begin{array}{ccccc}
 2     &-1&   0   & \ldots &0        \\
-1     & 2&  -1   & \ldots &0        \\
0      &-1&   2   & \ldots &0        \\
\vdots &  &       &\ddots  &\vdots   \\
0      &  &\ldots &        & 2 
\end{array}
\right| = r+1 \nonumber \\
\mbox{det}\A_0&=&n.
\label{trd1}
\end{eqnarray}

{\bf Mutually nonlocal branes with} $\mbox{Tr} K=2$. In this case the
overall monodromy has an eigenvector $\vect{p}{q}$ with eigenvalue
1. This means that a $\vect{p}{q}$ string can wind around the
7-branes.  The $\vect{p}{q}$-loop is a nontrivial junction whose
intersection with any elements of $\Lambda_0$ is zero. Choosing this
junction as one of the basis elements shows 
\begin{eqnarray}
\mbox{det}\A_0=0.
\label{trd2}
\end{eqnarray}

{\bf Mutually nonlocal branes with} $\mbox{Tr} K\neq2$ It is hard to
find an explicit basis of $\Lambda_0$ in general, but it is possible
to compute the volume of its unit cell without having one. To this
end, consider the following basis on $\Lambda$:
\begin{equation}
\Lambda = \{\alpha_i\}_{i=1}^{r}\cup\{{\bf j}_p,{\bf j}_q\},
\end{equation}
where $\{\alpha_i\}_{i=1}^{r}$ is a basis on $\Lambda_0$ while ${\bf
j}_p$ and ${\bf j}_q$ correspond to (proper) junctions carrying
asymptotic charges $\vect{1}{0}$ and $\vect{0}{\mQ}$,
respectively. (Such a basis exists since any basis of a sublattice can
be completed to the full lattice. First introduce ${\bf j}_p$
extending the basis on $\Lambda_0$ to a basis on the lattice of
junctions with zero $q$-charge, and then define ${\bf j}_q$ to
complete the basis on $\Lambda$.) Notice that when $\mbox{Tr} K\neq
2$, ${\bf j}_p$ and ${\bf j}_q$ can be expressed as
\begin{eqnarray}
\label{intoj}
{\bf j}_p=\momega^p+\tilde\alpha_p, \qquad
{\bf j}_q=\mQ\momega^q+\tilde\alpha_q
\end{eqnarray}
where $\momega^p$ and
$\momega^q$ as well as $\tilde\alpha_p$ and $\tilde\alpha_q$ are in
general improper junctions, i.e. lattice vectors with fractional
coefficients. As $\tilde\alpha_p$ and $\tilde\alpha_q$ lie in the
sublattice $\Lambda_0$, they can be expressed as:
\begin{eqnarray}
\tilde\alpha_p = \sum_{i=1}^{r} c_i\alpha_i ;\hspace{.5in}
\tilde\alpha_q = \sum_{i=1}^{r} d_i\alpha_i.
\label{tildedef}
\end{eqnarray}
The volume of the unit cell of $\Lambda$ in this basis can be computed
from the determinant:
\begin{eqnarray}
\mbox{det}\A=\left|
\begin{array}{ccccc}
-(\alpha_1,\alpha_1)&\ldots&-(\alpha_1,\alpha_r) &
-(\alpha_1,{\bf j}_p)& -(\alpha_1,{\bf j}_q)\\
\vdots&\ddots&\vdots&\vdots&\vdots    \\
-(\alpha_r,\alpha_1)&\ldots&-(\alpha_r,\alpha_r)&
-(\alpha_r,{\bf j}_p)& -(\alpha_r,{\bf j}_q) \\
-({\bf j}_p,\alpha_1)&\ldots&-({\bf j}_p,\alpha_r)&
-({\bf j}_p,{\bf j}_p)&-({\bf j}_p,{\bf j}_q)\\
-({\bf j}_q,\alpha_1)&\ldots&-({\bf j}_q,\alpha_r)&
-({\bf j}_q,{\bf j}_p)&-({\bf j}_q,{\bf j}_q)
\end{array}
\right|.
\end{eqnarray}
To evaluate, let us add the linear combination $-\sum_{i=1}^{r}
c_i\cdot\mbox{row}_i$ of the first $r$ rows to the $(n-1)$-th and
$-\sum_{i=1}^{r} d_i\cdot\mbox{row}_i$ to the $n$-th, and then do
similarly with the columns. Using
$(\alpha_i,\momega^p)=(\alpha_i,\momega^p)=0$ and \myref{tildedef}, we
obtain
\begin{eqnarray}
\mbox{det}\A=
\left|
\begin{array}{ccccc}
-(\alpha_1,\alpha_1)&\ldots&-(\alpha_1,\alpha_r) &0&0\\
\vdots&\ddots&\vdots&\vdots&\vdots    \\
-(\alpha_r,\alpha_1)&\ldots&-(\alpha_r,\alpha_r)&0&0 \\
0&\ldots&0& -(\momega^p,\momega^p)&-(\momega^p,\mQ\momega^q)\\
0&\ldots&0& -(\tilde\mQ\momega^q,\momega^p)&-(\mQ\momega^q,\mQ\momega^q)
\end{array}
\right|.
\end{eqnarray}

The 2-by-2 minor is straightforward to compute ($K \equiv ({a \; b
\atop c \; d } )$, $K-\unit$ is nonsingular):
\bea
\mbox{det}\A_{pq} \equiv \left|
\begin{array}{cc}
-(\momega^p,\momega^p) & -(\momega^p,\momega^q) \\
-(\momega^q,\momega^p) & -(\momega^q,\momega^q)
\end{array}\right| =
\left|\begin{array}{cc}
   \frac{c}{2-a-d} & \frac{d-a}{2(2-a-d)} \\
   \frac{d-a}{2(2-a-d)} & \frac{-b}{2-a-d} \end{array} \right|
= \frac{2+{\Tr} K}{4(2-{\Tr} K)},
\label{detpq}
\end{eqnarray}
where the elements of the matrix were identified using
\myref{fpqiu}, \myref{ffquad} and \myref{matform}. 
Given that  $\mbox{det}\A =
\mQ^2\mbox{det}\A_{pq}\;\mbox{det}\A_0$,
we can use \myref{detpq} and \myref{mystery} to find
\begin{eqnarray}
\mbox{det}\A_0 = \frac{2-{\Tr} K}{\mQ^2}
\label{trd3}
\end{eqnarray}
We thus conclude from \myref{trd1}, \myref{trd2}, \myref{trd3} that
every brane configuration satisfies the following trace-determinant
relation:
\begin{equation}
\begin{array}{|c|} \hline \\
\mQ^2\;\mbox{det}\A_0 = 2-{\Tr} K
\\ \\ \hline\end{array}
\label{mystery2}
\end{equation}
\myref{mystery2} is trivial for mutually local brane configurations
with $\mbox{Tr} K-2=0$ and $\mQ=0$, and the determinant is given by
\myref{trd1}. For other cases this equation determines the volume
$\sqrt{\mbox{det}\A_0}$ of the lattice of junctions with zero
asymptotic charge in terms of the monodromy of the configuration and
the invariant $\ell$.


\end{document}